\documentclass[modern]{aastex61}

\usepackage{bm}
\usepackage{graphicx}
\usepackage{textcomp}

\newcommand\<{$<$}
\usepackage{tablefootnote}
\submitjournal{ApJ}

\shorttitle{Dark Matter Distribution of Four Low-z Clusters of Galaxies}
\shortauthors{McCleary et al.}
\begin{document}
 
\title{Dark Matter Distribution of Four Low-$z$ Clusters of Galaxies}

\correspondingauthor{Jacqueline McCleary}
\email{Jacqueline\_McCleary@jpl.nasa.gov.}

\author[0000-0002-0786-7307]{Jacqueline McCleary}
\affil{Jet Propulsion Laboratory \\
California Institute of Technology\\
4800 Oak Grove Drive\\
Pasadena, CA 91109, USA}

\author{Ian dell'Antonio}
\affiliation{Brown University \\
184 Hope Street, Box 1843\\
Providence, RI 02912, USA}

\author{Anja von der Linden}
\affiliation{Stony Brook University \\
Dept. of Physics and Astronomy ESS 453\\
Stony Brook, NY 11794}



\begin{abstract}
We present here the weak gravitational lensing detection of four nearby galaxy clusters in the southern sky: Abell 2029, Abell 85, Abell 1606 and Abell 2457. The weak lensing detections of Abell 1606 and Abell 2457 are the first in the literature. This work capitalizes on the wide field of view of the Dark Energy Camera at the Cerro Tololo Inter-American Observatory, which we use to obtain deep, multi-wavelength imaging of all targets. We publish maps of the clusters' projected mass distributions, and obtain the $M_{200}$ of their clusters through NFW profile fits to the two-dimensional tangential ellipticity signal. 

\end{abstract}




%% USEFUL COMMANDS
%% \plotfiddle{PSFILE}{VSIZE}{ROTANG}{HSCALE}{VSCALE}{HTRANS}{VTRANS}
%% \plotfiddle{sample.eps}{2.6in}{-90.}{32.}{32.}{-250}{225}
%%

\section{Introduction} \label{sec:intro}

Galaxy clusters are the largest virialized structures in the
universe, and are a key to solving the problem of dark matter. %, as The shape and relative distributions of clusters' dark and baryonic matter strongly constrain allowed dark matter models. 
Offsets between the dark matter halo center and the brightest cluster galaxy probe the cross-section of self-interacting dark matter~\citep{2019MNRAS.488.1572H}. The shape of the cluster mass density profile also has a direct connection to models of warm or decaying dark matter~\citep{2014MNRAS.445..614W}. Moreover, because galaxy clusters virialize late in cosmic history, cluster number counts as a function of mass and redshift are powerful tests of dark energy. %% Can add citation if needed  DESC 2018

In this paper, we present 2-D projected mass maps for four nearby ($z < 0.10$) galaxy clusters observed with the Dark Energy Camera: Abell 2029, Abell 2457, Abell 85 and Abell 1606. We reconstruct cluster masses through their weak gravitational lensing signal: the distortion of background galaxy images by the foreground gravitational potential. This distortion manifests as a tangential alignment (or shear) of background galaxy images. By correlating the measured shapes of the background galaxies, a 2-D map of the projected cluster mass can be recovered. Weak lensing is well-suited to cluster studies, as it is a universal feature of galaxy cluster observations and measures the clusters’ mass distribution without needing to make assumptions about the composition or state of baryons. %To obtain an overall mass normalization for the maps, we fit parametric NFW models \cite{1997ApJ...490..493N} to the calibrated galaxy shear signal.  

There are several advantages to performing cluster weak lensing studies in the local $z\lesssim 0.10$ universe. Nearby clusters tend to be well-studied in the optical and X-rays, facilitating the study of the relative distributions of dark matter and baryons. Low redshift clusters subtend a large angle on the sky, making it easier to accurately measure the centroid and shapes of their mass density profiles. Member galaxies in these nearby clusters appear brighter, which makes it easier to create pure background galaxy catalogs (a problem for cosmological surveys of higher redshift clusters). Eventually, these factors will lead to tighter constraints on models of self-interacting dark matter than might otherwise be possible with higher-redshift cluster samples. Finally, all-sky X-ray and Sunya'ev-Zeldovich surveys are effectively complete at these low redshifts. Weak lensing maps like the ones presented in this study will robustly calibrate the relation between mass and observable (e.g., X-ray emissivity) at low redshift -- a crucial component of galaxy cluster-based dark energy studies.

The rest of this paper is organized as follows. Section \ref{sec:theory} summarizes weak lensing theory, and the cluster dataset is introduced in Section \ref{sec:obs}. In Sections \ref{sec:Methods1} and \ref{sec:Methods2}, the data reduction and catalog creation are discussed, including the PSF correction scheme. Our methods for weak lensing analysis and mass normalization are presented in Section \ref{sec:Methods3}. The results of our analysis are presented in Section \ref{sec:Results}, and we conclude with future directions for our research in Section~\ref{sec:Discussion}.

\section{Theory}\label{sec:theory}

By deflecting and distorting the images of galaxies in their background, massive objects like clusters act as gravitational lenses. The convergence $\kappa$ is a scalar quantity equal to the Laplacian of the gravitational potential of the lens, and is represented by a weighted surface mass density $\Sigma$:
\begin{equation}
\kappa \equiv \frac{1}{2} \nabla^2 \Psi(\theta)=\frac{\Sigma}{\Sigma_{\rm crit}};\quad \Sigma_{\rm crit} =\frac{c^2}{4\pi G}\frac{D_{\rm s}}{D_{\rm l} D_{\rm ls}}\label{eqn:sigmacrit}.
\end{equation}
The critical surface mass density $\Sigma_{\rm crit}$ of the lens depends on the angular diameter distances to the background galaxy $D_{\rm s}$, the lens $D_{\rm l}$ and $D_{\rm ls}$, respectively.

Observations of gravitational lenses return the {\em reduced shear} $\boldsymbol{g} =\frac{\boldsymbol{\gamma}}{1-\kappa}$. The convergence $\kappa$ produces an isotropic magnification of the galaxy image, while the shear $\gamma$ produces a curl-free stretching in the direction tangential to the lens. Areas of $\kappa \ll 1$ define the weak lensing (WL) regime, in which the distortion of background galaxy images produced by the lens is much smaller than the galaxy images themselves. In the weak lensing regime, the reduced shear measured on galaxy images is an unbiased estimator for the projected mass density of Equation~\ref{eqn:sigmacrit}. For a comprehensive treatment of weak lensing theory, see reviews by \cite{2001PhR...340..291B} and \cite{2002LNP...608...55W}.

Because the lensing potential induces curl-free distortions in galaxy images, we estimate the reduced shear with the {\em tangential ellipticity}:
\begin{equation}
e_{\tan} = -(e_1 \cos(2\phi) + e_2 \sin(2\phi)) \simeq 2\gamma \label{eqn:etan}.
\end{equation}
 The variables $e_1$ and $e_2$ in Equation~\ref{eqn:etan} are the polarization states of background galaxies with complex ellipticities $\boldsymbol{e}$; $\phi$ is the azimuthal angle from the fiducial center of mass to the galaxy. 
n the absence of a gravitational lens (and spurious ellipticity from the PSF), the azimuthally averaged $\langle e_{\tan}\rangle$ vanishes. Hence, the $\langle e_{\tan}\rangle$ is an unbiased estimator for the WL shear $\gamma$ at a location in the observation. 

Because it is a curl-free statistic, in analogy with electromagnetism, Equation~\ref{eqn:etan} is sometimes called E-mode signal. A divergence-free statistic, the B-mode, is obtained by rotating Equation~\ref{eqn:etan} through $\pi/4$ radians:
\begin{equation}
e_{\rm c} =e_2\cos(2\phi) - e_1\sin(2\phi)\label{eqn:gc}.
\end{equation}

Since most systematics are expected to add equal power to E- and B-modes \citep{2003AJ....125.1014J}, B-mode maps generated with $e_{\rm c}$ probe systematic errors in our analysis. 

Galaxy shapes are convolved with the point spread function (PSF) of the telescope and atmosphere. The PSF circularizes the objects (thereby diluting the weak lensing signal) and induces ellipticities into the galaxy shapes that can mimic WL signal. To recover pre-seeing shapes and an unbiased $e_{\rm tan}$, we use the KSB algorithm developed in \cite{1995ApJ...449..460K}, \cite{1997ApJ...475...20L} and \cite{1998ApJ...504..636H}, and extended by \cite{2001A&A...366..717E}. In this scheme, the observed ellipticity $\boldsymbol{e^{\rm obs}}$ of a galaxy is the sum of three components:

\begin{equation}
\boldsymbol{e^{\rm obs}} = \boldsymbol{\hat{e}}^0+P^{\rm g}\boldsymbol{g}+P^{\rm sm}\frac{\boldsymbol{e}^{\rm \star obs}}{P^{\rm \star sm}};\quad P^{\rm g}=P^{\rm sh}-P^{\rm sm}(P^{\rm \star sm})^{\rm -1}P^{\rm \star sh}\label{eqn:KSBellip}
\end{equation}
The galaxy's intrinsic ellipticity is represented as $\boldsymbol{\hat{e}}^0$. The ``pre-seeing'' shear polarizability tensor $P^{\rm g}$ contains a a correction for the (isotropic) circularization induced by atmospheric seeing and the shear polarizability tensor $P^{\rm sh}$, which describes the galaxies' susceptibility to astrophysical shear. The stellar anisotropy kernel $\boldsymbol{e^{\rm \star obs}}/P^{\rm \star sm}$ describes the anisotropic part of the PSF, and is measured from the ellipticities of observed stars in the observation. The smear polarizability tensor P$^{\rm sm}$ characterizes the susceptibility of objects to PSF anisotropy, and depends largely on the object size. 
Averaged over many background galaxies with no intrinsic alignment, the KSB algorithm returns the reduced shear:

\begin{equation}
\boldsymbol{\hat{g}} = (P^{\rm g})^{-1}\boldsymbol{e}^{\text{aniso}};\quad \boldsymbol{e}^{\text{aniso}}= \boldsymbol{e^{\rm obs}}-P^{\rm sm}\frac{\boldsymbol{e}^{\rm \star obs}}{P^{\rm \star sm}}\label{eqn:ghat}
\end{equation}

Since in general the off-diagonal part of the $P^g$ tensor is much smaller than the trace, the following approximations are made:
\begin{equation}
(P^{\rm \star sm})^{\rm -1}P^{\rm \star sh} \rightarrow \frac{{\rm Tr}[P^{\rm \star sh}]}{{\rm Tr}[P^{\rm \star sm}]}\equiv T^\star;\quad (P^{\rm g})^{-1} \rightarrow \frac{2}{{\rm Tr}[P^{\rm g}]}
\end{equation}
These approximations have the effect of simplifying calculations and also reducing sensitivity to noise \citep{2001A&A...366..717E,2006MNRAS.368.1323H}. The $e_1$ and $e_2$ of Equation~\ref{eqn:etan} are then replaced by the equivalent polarization states of $\hat{g}$.

After PSF correction, we identify shear peaks using the aperture mass statistic $M_{\rm ap}$ \citep{1996MNRAS.283..837S}. For discrete background sources, the aperture mass statistic has the form

\begin{equation}
M_{\rm ap}(\theta_0) = \frac{1}{n}\sum_i^{N_{gals}} e^{\tan}_i(\theta)Q(|\theta_0 - \theta|),\label{eqn:aperturemassstat}
\end{equation}
where the sum is taken over all galaxies in the observation and $n$ is the number density of galaxies in the image. Formally, $Q(|\theta_0 - \theta])$ is a weight function that maximizes the S/N of the observation over some characteristic scale $\theta_0$ and vanishes on a scale larger than the filter's ``aperture.'' By design, the $M_{\rm ap}$ is a local measurement involving only the shear from galaxies within an angle $\theta_0$ of the center at position $\theta$.

In this work, we use an approximate Weiner filter for NFW halos in the presence of large-scale structure ``noise'' \citep{2004A&A...420...75S} in calculations of aperture mass. The filter is given as  
\begin{equation}
Q(x) = \frac{1}{(1+e^{a-bx} + e^{dx-c})}\frac{\tanh(x/x_c)}{\pi R_{\rm S}^2(x/x_c)} \label{eqn:Schirmerfilter},
\end{equation}
where $R_{\rm S}$ is the filter radius and $x=r/R_{\rm S}$ is a scaled distance between the cluster center and the point in consideration. To optimize this so-called Schirmer filter for detection of NFW shear profiles, the parameters in Equation \ref{eqn:Schirmerfilter} are tuned to $a=6, b=150, c=47, d=50$ and $x_c=0.12$ \citep{2005A&A...442...43H}. Noting that the Schirmer filter weights peak sharply at a value of $x_cR_{\rm S}$, the structures identified have characteristic size $\sim 0.12 R_{\rm S}$.

\section{Cluster Sample} \label{sec:obs} 
All observations were taken with the Dark Energy Camera (DECam) at the Cerro Tololo Inter-American Observatory's 4-meter telescope. The DECam imager consists of 62 2048 $\times$ 4096 pixel science CCDs (60 of which are currently operational) arranged in a hexagon, and captures 2.2 square degrees at 0\farcs265/pixel scale in one exposure\citep{2008SPIE.7014E..0ED, 2015AJ....150..150F}. In the rest frame of our average cluster redshift of $z=0.06$, the camera spans an area 9.2 Mpc wide. With this field of view, DECam allows us to image the entire virial region of a low-redshift cluster in a single pointing, making the instrument a natural choice for our project. 

Clusters in this project are drawn from two separate observing programs: a dedicated campaign by JM, and a DECam program by AvdL to obtain scaling relations for cluster cosmology. As a consequence, the data were taken under a range of seeing conditions, a situation for which we control in our analysis. In addition, the weak lensing shape measurement (see Section~\ref{sec:Methods3}) is carried out in two different wavelengths: clusters observed by AvdL have their shape measurement performed in DECam i band, while those clusters observed by JM have their deepest data in r. Clusters were observed in r (or i) when the seeing FWHM reached \<~1\arcsec~and in ugi(r)z otherwise. Accordingly, shape analysis-quality imaging has uniformly good resolution, as well as a greater depth than the imaging in other filters. Data in non-shape analysis filters are used to provide color information for photometric redshifts (see Section~\ref{sec:Methods2}). 

\begin{deluxetable*}{cCCCcc}[bt]
\tablecaption{\footnotesize{Clusters Analyzed in this Study}\label{tab:sample}}
\tablecolumns{6}
\tablewidth{0.99\textwidth}
\tabletypesize{\scriptsize}
\tablehead{
\colhead{Name} &
\colhead{Redshift}&
\colhead{$\alpha$} &
\colhead{$\delta$} & \colhead{Shape Analysis Filter} & \colhead{Photometry Filters} \\
\colhead{} & \colhead{(z)} &\colhead{J2000.0} & \colhead{J2000.0} & \colhead{\& Total Exposure Time} & \colhead{\& Total Exposure Time}
}
\startdata
Abell 2029 & 0.0774 & 15^h 10^m 58\fs7 &+05\arcdeg 45\arcmin 42\arcsec & i (4920 s) & u (2200 s) g (3150 s) r (3700 s) z (5900 s) \\
Abell 1606 &0.0963& 12^h 44^m 36\fs4 & -11\arcdeg 59\arcmin 24\arcsec & i ( 5890 s) & u (1900 s) g (2200 s) r (2760 s) z (3400 s) \\
Abell 85 & 0.0557 & 00^h 41^m 37\fs8 &  -09\arcdeg 20\arcmin 33\arcsec & r (3530 s) & u (6630 s) g (2900 s) i (1500 s) z (1330 s) \\
Abell 2457 & 0.0591  & 22^h 35^m 40\fs3 & +01\arcdeg 31\arcmin 34\arcsec & r (3230 s) & u (6740 s) g (4200 s) i(1500 s) z (4200 s) \\
\enddata
\end{deluxetable*}

At present, the full sample comprises 11 Abell clusters for which all required observations are complete. Member clusters were selected for X-ray luminosity greater than $L_X > 10^{44}$ ergs (a proxy for high mass), $z \lesssim 0.12$\footnote{We choose this cutoff for for the low-redshift sample, as opposed to $z=0.10$ or $z=0.14$, because $z<0.12$ is the completion threshold for clusters of $L_X=10^{44}$ ergs in the flux-limited RASS survey} and existing X-ray data sufficiently deep to allow comparison of dark matter overdensities and the hot cluster gas. In this phase of the study, we restrict ourselves to those clusters overlapping with the SDSS DR9 footprint, to facilitate photometric calibration of the images used for the photo-z determination: Abell 2457, Abell 1606, Abell 85 and Abell 2029.  Observation information for these four clusters is summarized in Table~\ref{tab:sample}. 

\section{Methods: Catalog Creation, Calibration and Cuts} \label{sec:Methodsme}
\subsection{Methods: Data Reduction} \label{sec:Methods1}
The NOAO has made available the DECam Community Pipeline (CP), an automatic, high performance processing system which applies the best instrumental calibrations available at the time the data is collected. The CP includes: bias calibration; crosstalk; masking and interpolation over saturated and bad pixels; CCD non-linearity and the flat field gain calibration; fringe pattern subtraction; astrometric calibration; single exposure cosmic ray masking; characterization of photometric quality; sky pattern removal; and illumination correction. In addition to sky images, the CP produces inverse variance weight maps for DECam science exposures. These contain information on e.g.~transient objects or bad pixels that should not be included in the final stacked image. For full descriptions of the DECam pipeline processing system, see chapter 4 of the NOAO Data Handbook~\citep{NOAODHB2.2}.

Reprojection of CCD image sub-sections is accomplished with {\sc SWarp}. The combination of images is performed using a clipped mean extension, which is exceptionally stable to a wide range of artifacts in individual frames and produces a stacked image whose point-spread function (PSF) is a linear combination of the single frame PSFs \citep{2014PASP..126..158G}. To avoid degrading the final stacked images, any exposures with a stellar FWHM greater than $\sim 1\farcs75$ are excluded. We also create for each cluster a lensing-quality stack only from CCD images with stellar FWHM less than 1\arcsec. Shape measurement is based on the lensing-quality stacks. 

\subsection{Catalog creation and filtering}\label{sec:Methods2}

As in JM15, background source catalogs from our cluster image stacks are generated with {\sc SExtractor}. \textbf{We characterize source galaxies with aperture magnitudes of 15 pixels in diameter, and run {\sc SExtractor} in dual-image mode}. To avoid introducing color gradients within an object, image stacks in all bandpasses are convolved with a Gaussian filter that degrades the stellar FWHM to match the  worst stellar FWHM in the set (usually u). However, if the seeing differences are too large, the Gaussian scaling of the PSF is expected to break down. We therefore adopt the strategy of Weighing the Giants, and limit the maximum PSF size of any included images to no more than the seeing of the detection image plus $0\farcs 3$. 

Low {\sc SExtractor} significance and deblending thresholds yield highly complete catalogs of galaxies, but also a fair number of spurious detections. These ``objects'' are filtered out with a number of quality cuts common to all weak lensing analyses. For every cluster, we filter out objects fainter than the 50\% completeness limit in that cluster's lensing band: $r=24.3$ for Abell 2457, $i=23.9$ for A1606, $r=24.2$ for A85 and $i=24.4$ for Abell 2029. The error bar on all limits is $\pm 0.02$ magnitudes. We also filter out both stars and small, poorly-measured objects though a requirement that objects be 15\% larger than the size of the stellar PSF ({\sc Analyseldac} half-light radius \texttt{r$_{\rm h}$} $\gtrsim 2.1$, see Section~\ref{sec:analyseldac}).

\subsection{Photometric redshift fitting} \label{Methods2:BPZ}
The images of galaxies in the foreground of clusters (and the cluster galaxies themselves) do not experience shear from gravitational lensing; their presence in the source catalog dilutes the measured aperture mass and they should therefore be filtered out. In addition, the angular diameter distances to the sources $D_{\rm S}$ and between the sources and lens $D_{\rm LS}$ must be known to obtain a mass normalizations for the cluster aperture mass maps (See Eq.~\ref{eqn:sigmacrit}). To filter out low-redshift contaminants and obtain angular diameter distances, galaxy redshifts are obtained using the Bayesian photometric redshift software BPZ~\citep{2000ApJ...536..571B, 2006AJ....132..926C} with the standard HDFN prior, the SWIRE template library~\citep{2007ApJ...663...81P} with eight levels of interpolation between neighboring templates, and probability spikes at the cluster redshifts. Redshifts are considered over the range $0.005 < z_{\rm BPZ} < 3.0$.
% Figure 1
\begin{figure}[htb]
\begin{center}
\includegraphics [width=0.8\textwidth]{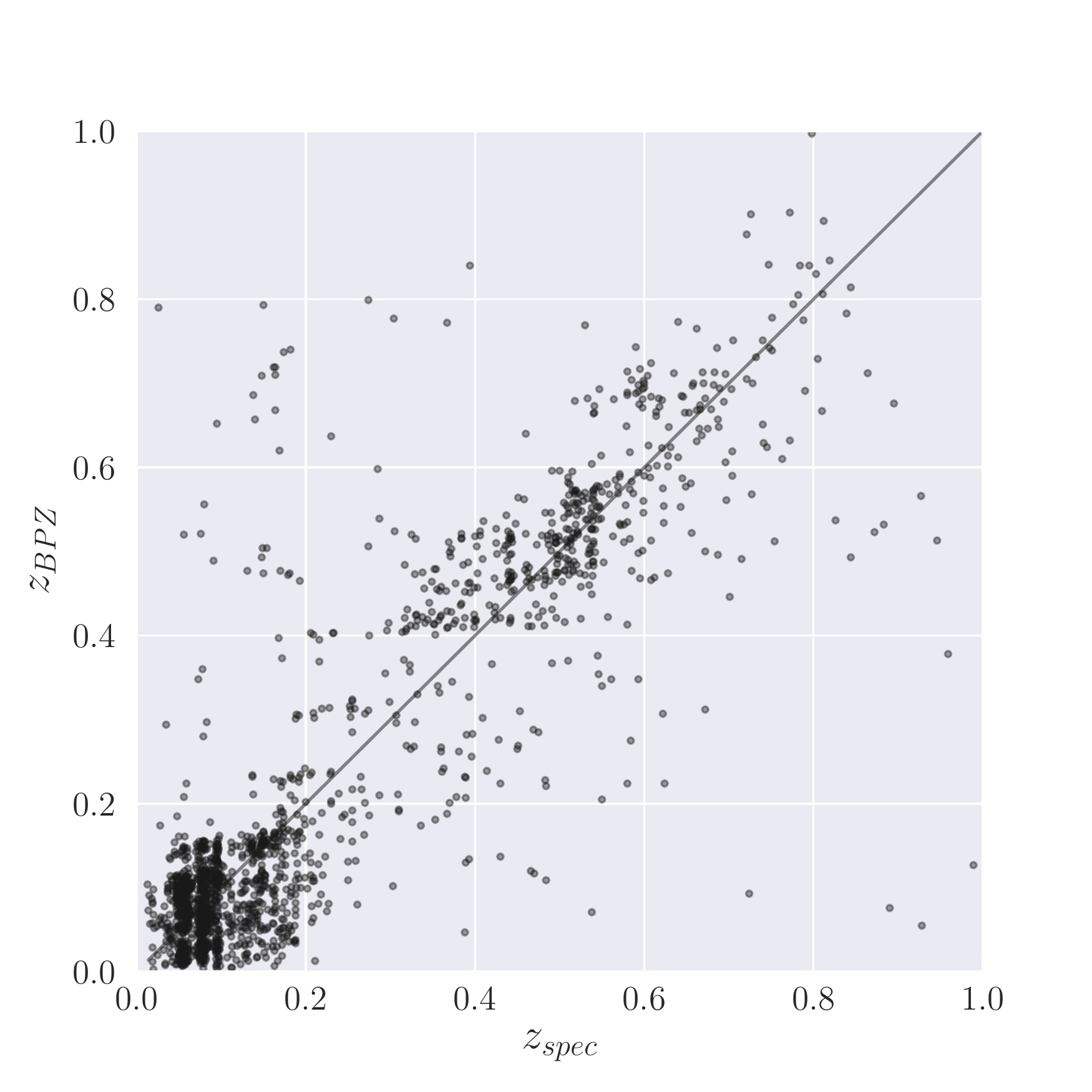}
\caption{\small{Best-fit BPZ redshift plotted against BPZ redshift for all clusters.}}
\label{fig:zbzspec}
\end{center}
\end{figure}

Robust photometric redshifts require accurate photometric calibration. Training sets of galaxies with spectroscopic redshifts are used to calibrate zeropoint offsets in each filter; spectroscopic redshifts are obtained with the NASA/IPAC Extragalactic Database\footnote{The NASA/IPAC Extragalactic Database (NED) is operated by the Jet Propulsion Laboratory, California Institute of Technology, under contract with the National Aeronautics and Space Administration.}. Redshifts for A85 come from~\cite{2016MNRAS.458.1590A}. For A2029, redshifts of 1,215 galaxies are obtained from \cite{2018arXiv180901137S}. Redshifts for A2457 are taken  from~\cite{2015A&A...581A..41G}. For A1606, we use the redshifts published in~\cite{2000ApJS..130..237T}. A scatter plot of spectroscopic vs. photometric redshifts is shown in Figure~\ref{fig:zbzspec}. The median $\Delta z=(z_{\rm BPZ}-z_{\rm spec})$ scatters around $0.001$ but with a high $\sigma_{\rm \Delta z} \sim 0.21$. This range is larger than the per-galaxy RMS error found by e.g., \citealp{2014MNRAS.439...28K}. Our larger error bars may be attributed to the very low redshifts of the cluster members that make up a disproportionate number of the spectroscopic sample, and the known difficulties of BPZ with low-redshift clusters. Provision is made for the uncertainty in $\Delta z$ in the creation of a background galaxy sample by using the BPZ posterior probability distributions. Galaxies promoted to analysis in Section \ref{sec:ConvergenceMapping} are required to have less than a 20\% probability of being at a redshift below the cluster redshift, plus a margin of $0.1$: $P(z_{\rm BPZ} <z_{\rm clust}+0.10) \leq 20\%$. We found that this method of background galaxy selection yields higher S/N aperture mass maps than when a background sample based on a single-point redshift cutoff is used. %This technique was first implemented in \cite{2014MNRAS.439...48A}. They are not actually the first to use this, I think. 

\section{Methods: Weak Lensing Analysis }\label{sec:Methods3}
\subsection {Shape Measurement, PSF correction and STEP calibration}\label{sec:analyseldac}
%Figure 1
\begin{figure}[ht]
\begin{center}
\includegraphics[width=0.6\textwidth,trim={0 4.5cm 0 4.5cm},clip]{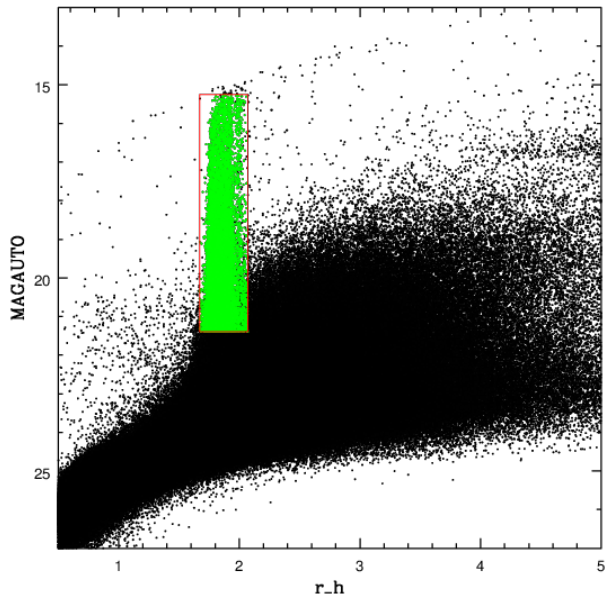}
\caption{{\small Diagram of size (x-axis, `\texttt{r$_{\rm h}$'} versus magnitude (y-axis, `\texttt{MAGAUTO}') from the source catalog of Abell 85. Plotted size \texttt{r$_{\rm h}$} is the {\sc Analyseldac} half-light radius, and magnitudes are the {\sc SExtractor} \texttt{MAGAUTO}, designed give the most precise estimate of ``total magnitudes'' for faint objects. The green box marks stars selected for use in the KSB PSF correction.}}
\label{fig:sizemag}
\end{center}
\end{figure}

Telescope optics induce anisotropy in the PSF of observed objects anisotropy in the PSF of observed objects (the $P^{\rm sm}$ tensor in Eq.~\ref{eqn:KSBellip}), making their shapes locally correlated and mimicking WL shear signal. PSFs on telescopes like DECam are $2-4\%$ elliptical, dwarfing the lensing signal from the cluster. The PSF has an additional isotropic component from atmospheric ``seeing,'', which circularizes object shapes and dilutes the weak lensing signal. The removal of the PSF from the images of observed galaxies is thus crucial to the success of weak lensing analyses.  

We adopt the KSB algorithm for PSF correction, which simulations such as STEP2~\citep{2007MNRAS.376...13M} have shown to perform well in the low-shear regime. The KSB algorithm assumes that the PSF can be described as the convolution of a compact anisotropic kernel and a large isotropic kernel, and the correction is applied at the catalog level (rather than convolved directly with telescope images). To facilitate this stage of our analysis, members of the Weighing the Giants team shared the pipeline for the KSB implementation described in~\cite{2014MNRAS.439...28K} We describe our application of the Weighing the Giants shape measurement pipeline; WTG1 Sections $5.1-5.6$ and references therein contain a complete discussion of the software. 

SExtractor shape catalogs and images are supplied to the {\sc Analyselac} code, which returns the second intensity moments and tensor components of sources in the observations. In the limit of a perfectly isotropic PSF, stars are perfectly round ($|e|\sim 0$), so the PSF correction is determined from a sample of bright but unsaturated stars, which is identified from the size-magnitude diagram of Figure~\ref{fig:sizemag}. The region highlighted in green in Figure~\ref{fig:sizemag} reflects a balance between keeping as many stars as possible to cover the entire field of view and a clean sample of stars to avoid circularizing away the ellipticity signal of small circular galaxies in the region where the stellar locus merges into the galaxy distribution. 

\begin{figure}[htp]
\begin{center}
\includegraphics[scale=0.8]{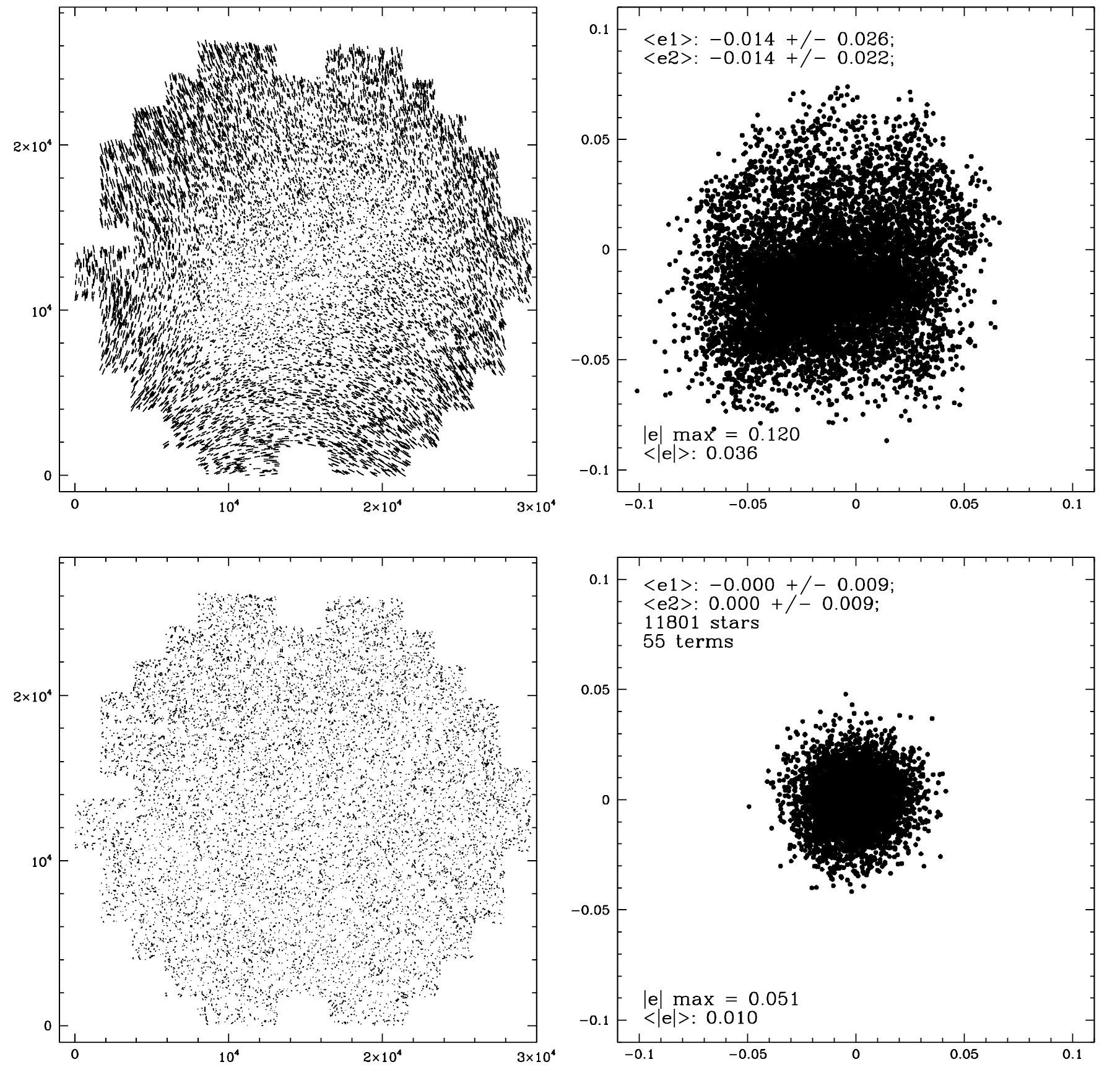}
\caption{\small{ Illustration of the PSF correction applied to the Abell 85 stars highlighted in green in Figure~\ref{fig:sizemag}. The top-left panel shows the uncorrected stellar ellipticity pattern, traced out by lines at the position of each star. The line lengths are proportional to the magnitude of the ellipticity $|e^\star|$ and line orientations are equal to $\phi=\frac{1}{2} \arctan(e^\star_2/ e^\star_1)$. The top-right panel shows the distribution of uncorrected $e^\star_1$ and $e^\star_2$ values. The bottom-left panel shows the residuals in the stellar ellipticity pattern after correction with a 9th-order polynomial; the distribution of corrected $e^\star_1$ and $e^\star_2$ values is shown in the bottom-right panel. The mean stellar ellipticity has been reduced from $3.6\%$ at the edges of the field to 1\%.}}
\label{fig:psfcorr}
\end{center}
\end{figure}
 The PSF anisotropy $P^{\rm sm}(\boldsymbol{e^{\rm \star obs}}/P^{\rm \star sm})$ is measured at the location of each star, and its variation across the field of view is interpolated using a polynomial model in x and y. As in WTG1, a 10-fold cross-validation procedure is used to determine the best order of polynomial fit. The stars used for PSF correction are first randomly subdivided into 10 groups. Each order of polynomial fit to the stellar ellipticity is recomputed with stars in nine of the ten groups, and residuals of the fit are computed on stars in the tenth group. In this way, ellipticity residuals are available for each star without actually using that star in the fit. The procedure repeats for all groups of stars, and for all polynomial orders. The polynomial order that minimizes the sum of the standard deviation of the two ellipticity components $e^\star_1$ and $e^\star_2$ is chosen as the best fit and applied to all objects in the catalog. Owing to the large size of the DECam field of view, the PSF variation in our catalogs was best captured by high-order polynomials (9th up to 12th order). Since there are of order 10,000 stars per cluster, this is a highly constrained problem despite the large number of degrees of freedom.  An
example of a successful solution is shown in Figure~\ref{fig:psfcorr}.

\begin{figure} [hbt] 
\begin{center}
\includegraphics[width=0.45\textwidth]{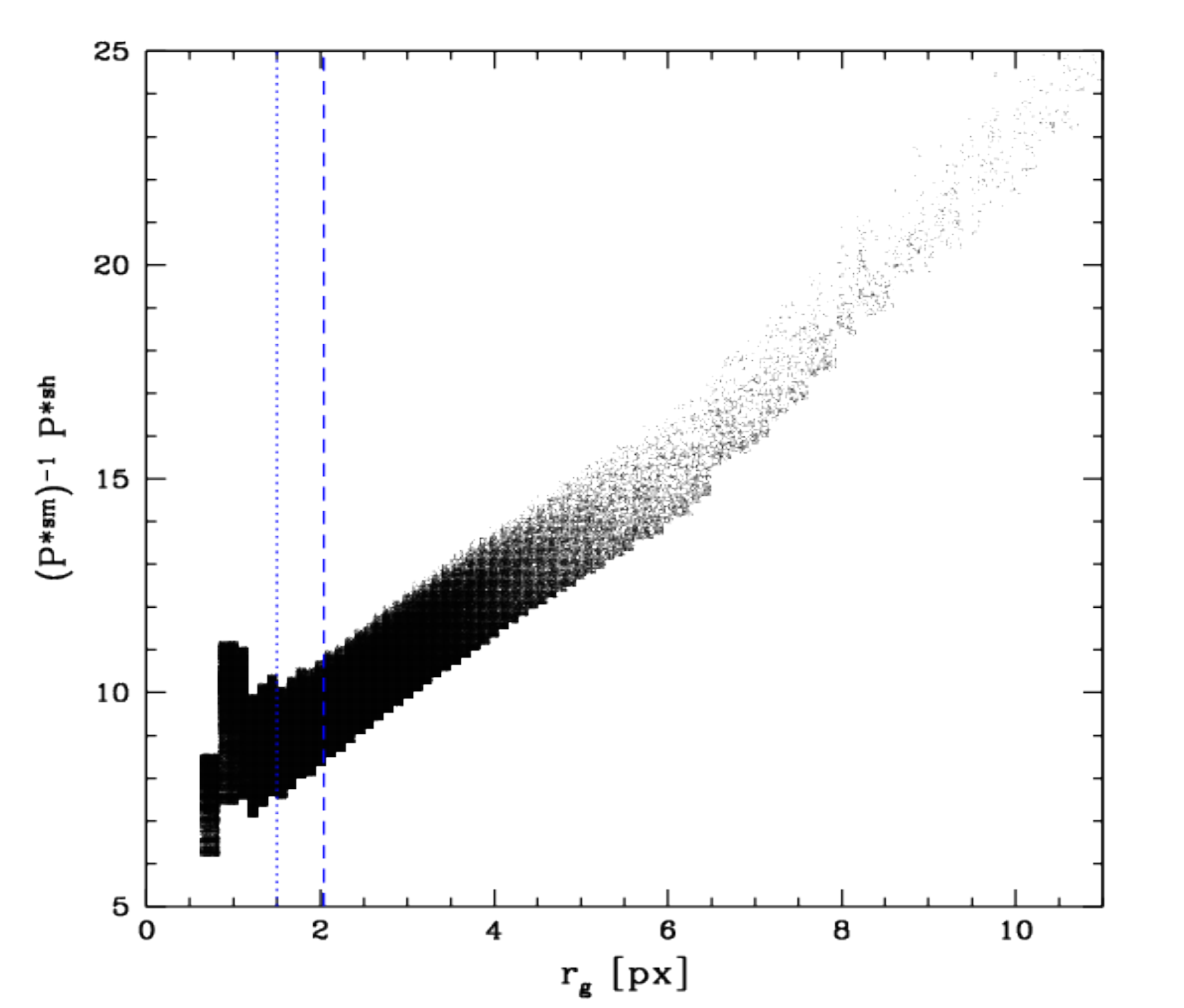}
\includegraphics[width=0.45\textwidth]{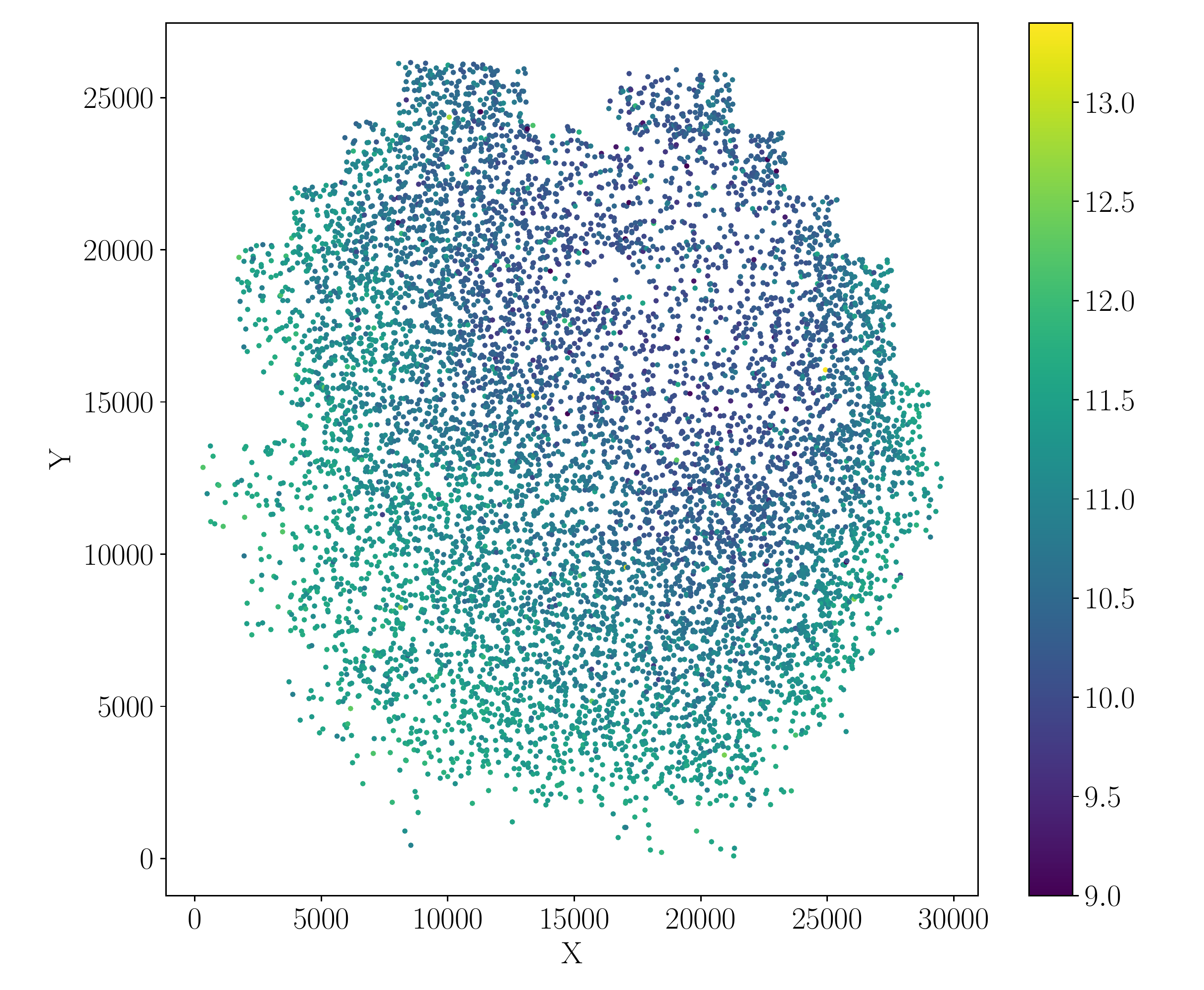}
\caption{\small{Illustration of the PSF isotropy correction as a function of position and object size for Abell 85. Left: $T^\star$ as a function of object size.The spread for a given object size reflects the spatial variation of $T^\star$ plotted on the left. The dashed blue line shows the mean \texttt{r$_{\rm g}$} of stars in the A85 observation, while the thin blue line shows the minimum size cutoff of 1.5 pixels for \texttt{r$_{\rm g}$}.} Right: variation of $T^\star=Tr[P^{ \rm \star sh}]/Tr[P^{ \rm \star sm}]$ across the DECam field of view. Each point marks the location of a star, and the color indicates the value of $T^\star$ when evaluated with a weight function of radius \texttt{r$_{\rm g}^{ \rm b}$}=3.1 pixels or 0\farcs82. }  
  \label{fig:isotropy}
  \end{center}
\end{figure}

After the anisotropic part of the PSF has been corrected, the isotropic part of the PSF ($P^{ \rm g}$ in Eq.~\ref{eqn:KSBellip}) may be determined by measuring $T^\star={\rm Tr}[P^{ \rm \star sh}]/{\rm Tr}[P^{ \rm \star sm}]$. The susceptibility of objects to the isotropic component of the PSF depends strongly on their size, which is expressed in the KSB formalism with a Gaussian weight function of width \texttt{r$_{\rm g}^{ \rm b}$}. Here, the weight \texttt{r$_{\rm g}^{ \rm b}$} is set to the objects' measured sizes $\texttt{r$_{\rm g}$}$. As the physical size of the PSF varies within the field of view of a telescope, $T^\star$ also varies spatially, independent of the object size. The left panel of Figure~\ref{fig:isotropy} shows the spatial variation of $T^\star$ for a representative value of the weight function \texttt{r$_{\rm g}^{ \rm b}$}.

Given the best-fit anisotropy polynomial, $T^\star$ is computed at discrete values of the weight function \texttt{r$^{\rm b}_{\rm g}$} over the range $0.33 \leq $ \texttt{r$_{\rm g}^{ \rm b}$} $\leq 18$ in 0.33-pixel increments.
At each bin in \texttt{r$^{ \rm b}_{\rm g}$}, we fit the spatial variation of $T^\star$ with a second-order polynomial, which suffices to capture its spatial variation. Each object in the catalog is then assigned a $T^\star$ based on the object's own size \texttt{r$_{\rm g}$} and position in the field of view. The right-hand panel of Figure~\ref{fig:isotropy} shows that $T^\star$ is roughly linear with size \texttt{r$_{\rm g}$} for objects significantly larger than the PSF (marked by the blue dotted line). Figure~\ref{fig:isotropy} shows significant pixelization artifacts for objects about the size of the PSF, which explains the size cut imposed in Section~\ref{sec:Methods2}. For comparison, the median \texttt{r$_{\rm g}$} value of stars in our catalog is shown as a dashed blue line.

Galaxies in the catalog must be corrected for the tendency of the KSB algorithm to underestimate shear, which will lead to an underestimate of the cluster masses~\citep{2001A&A...366..717E}. We use the procedure of WTG1 and \cite{2014MNRAS.439...48A}, themselves based on the simulations from the STEP2 Project~\citep{2007MNRAS.376...13M}, to calibrate ellipticities as a function of the S/N and size of each galaxy.

Once the anisotropic and isotropic parts of the PSF are computed for every object in the catalog, the reduced shear $\boldsymbol{\hat{g}}$ is given by Eq.~\ref{eqn:ghat}. 
 Although no upper size cut is applied to the catalogs, as the clusters in our samples are at very low redshifts, we apply a cut of $\boldsymbol{\hat{g}}<1.4$ before submitting galaxies to WL analysis as a control for unphysical PSF corrections. Only $\sim 10\%$ of objects larger than the PSF failed to meet this criterion. 
After all cuts have been applied, the final A2029 catalog has 210,206 objects; the A85 catalog has 197,456 objects; the A1606 catalog has 199,219 objects; and the A2457 catalog contains only 160,758. The corresponding background galaxy density ranges from 14 to 16 galaxies per square arcminute. 

\begin{figure}[p]
\begin{center}
\includegraphics[scale=0.45]{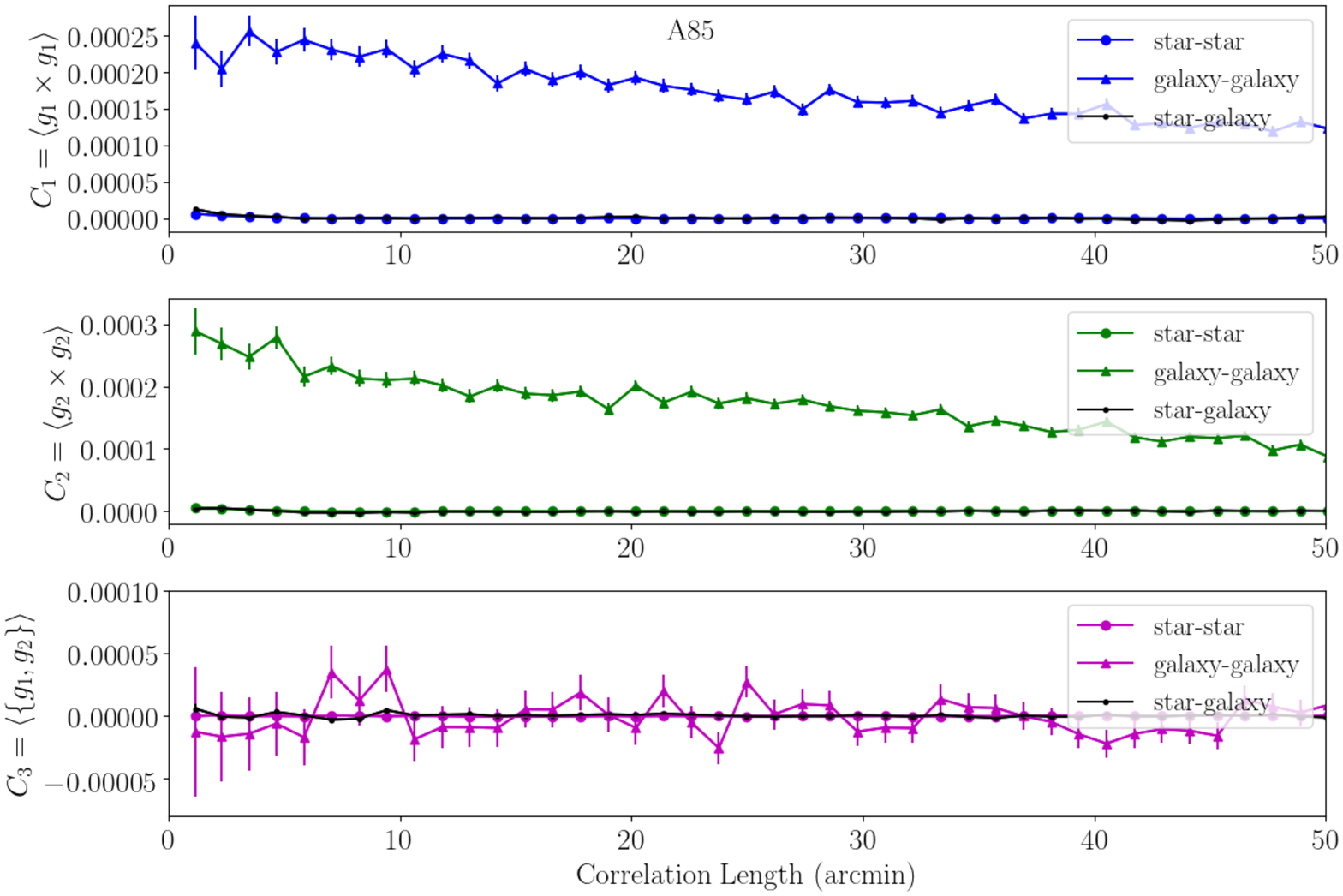}
\includegraphics[scale=0.45]{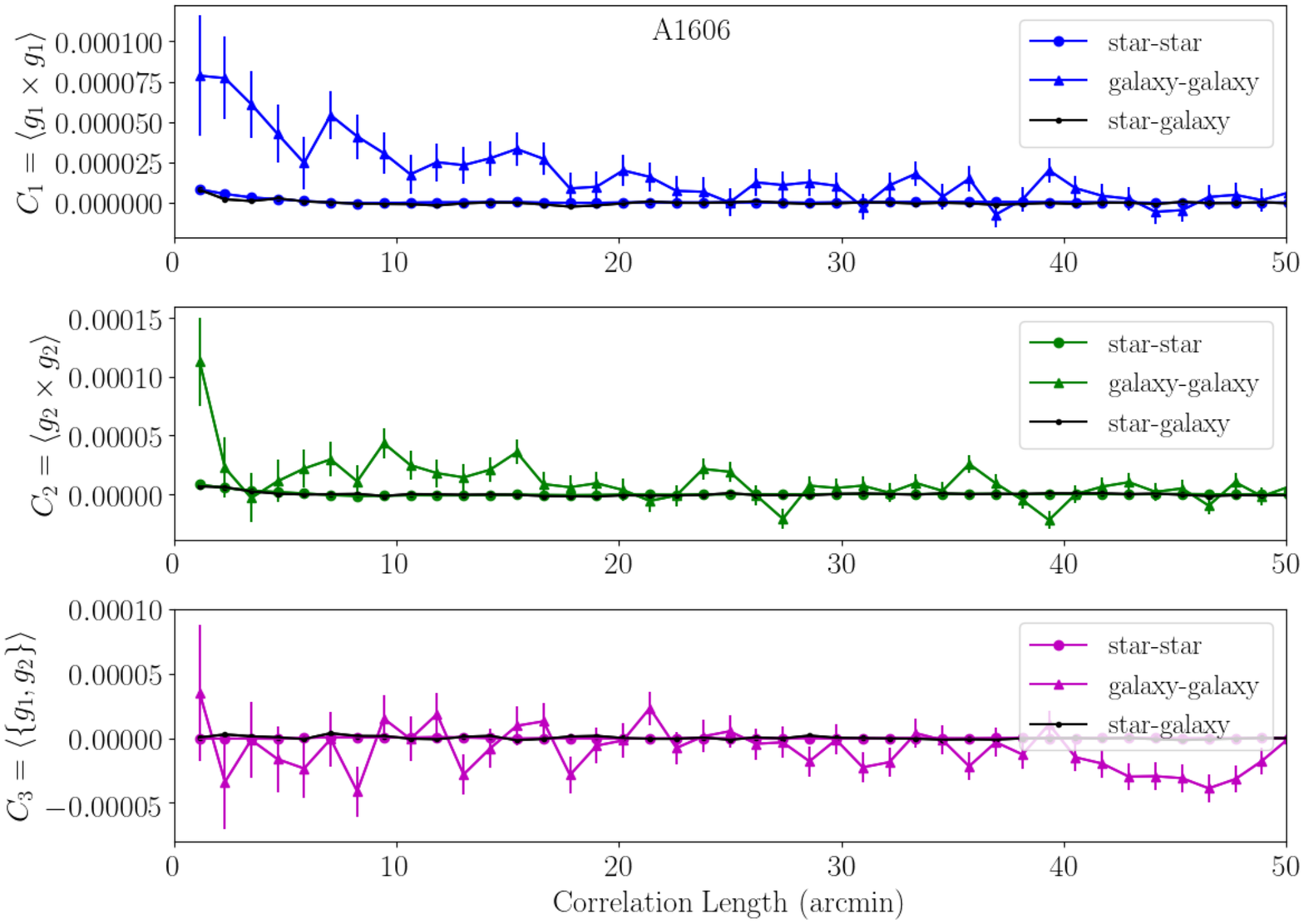}
\caption{{\small Correlations computed between ellipticity components for objects in the A85 observation (top) and A1606 observation (bottom). Units for all plots are ellipticity squared, modulated by the amplitude of the correlation. Error bars are the variance in each bin.}}
\label{fig:correlfuncs}
\end{center}
\end{figure}

\begin{figure}[htbp]
\begin{center}
\includegraphics[scale=0.45]{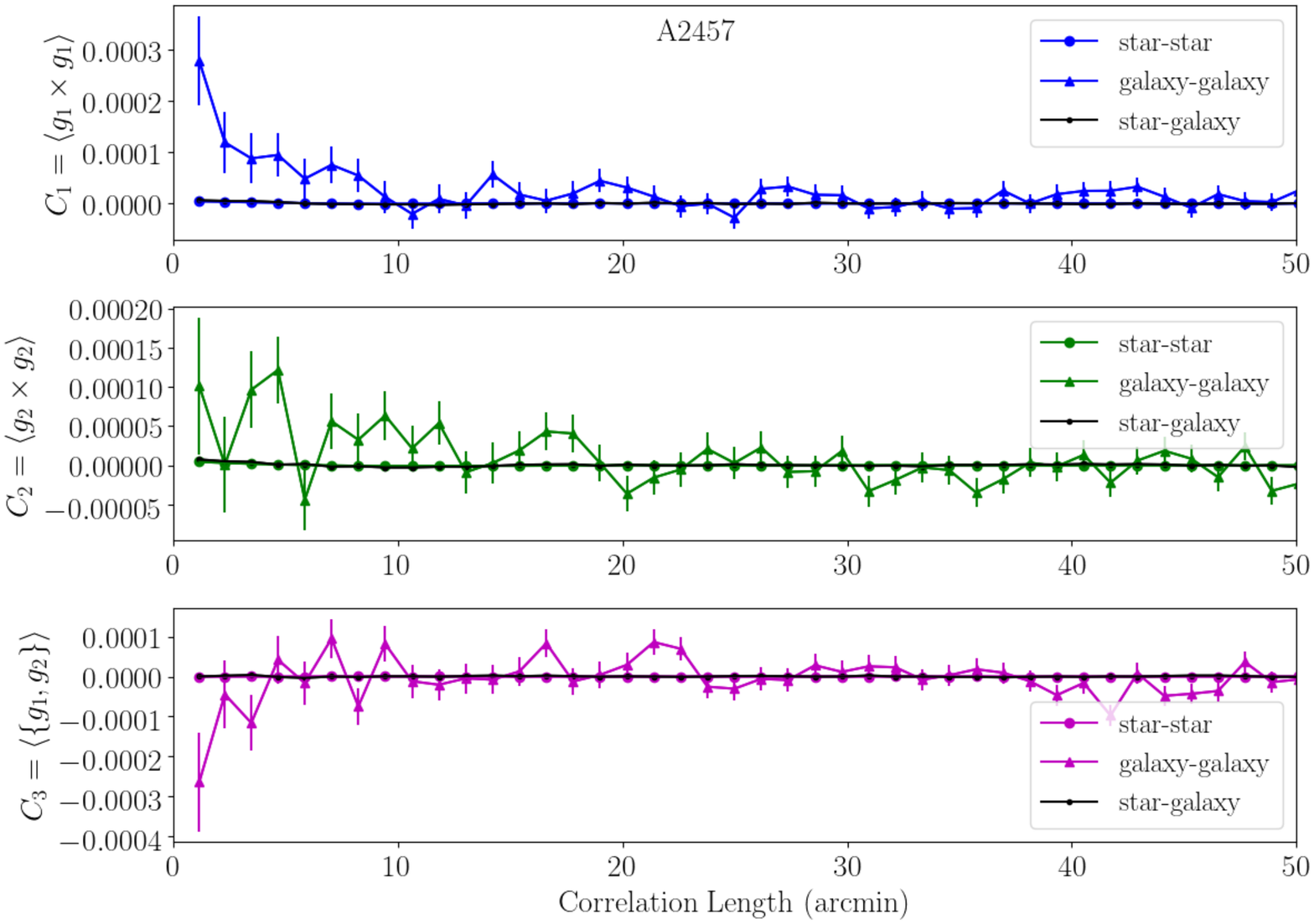}
\includegraphics[scale=0.45]{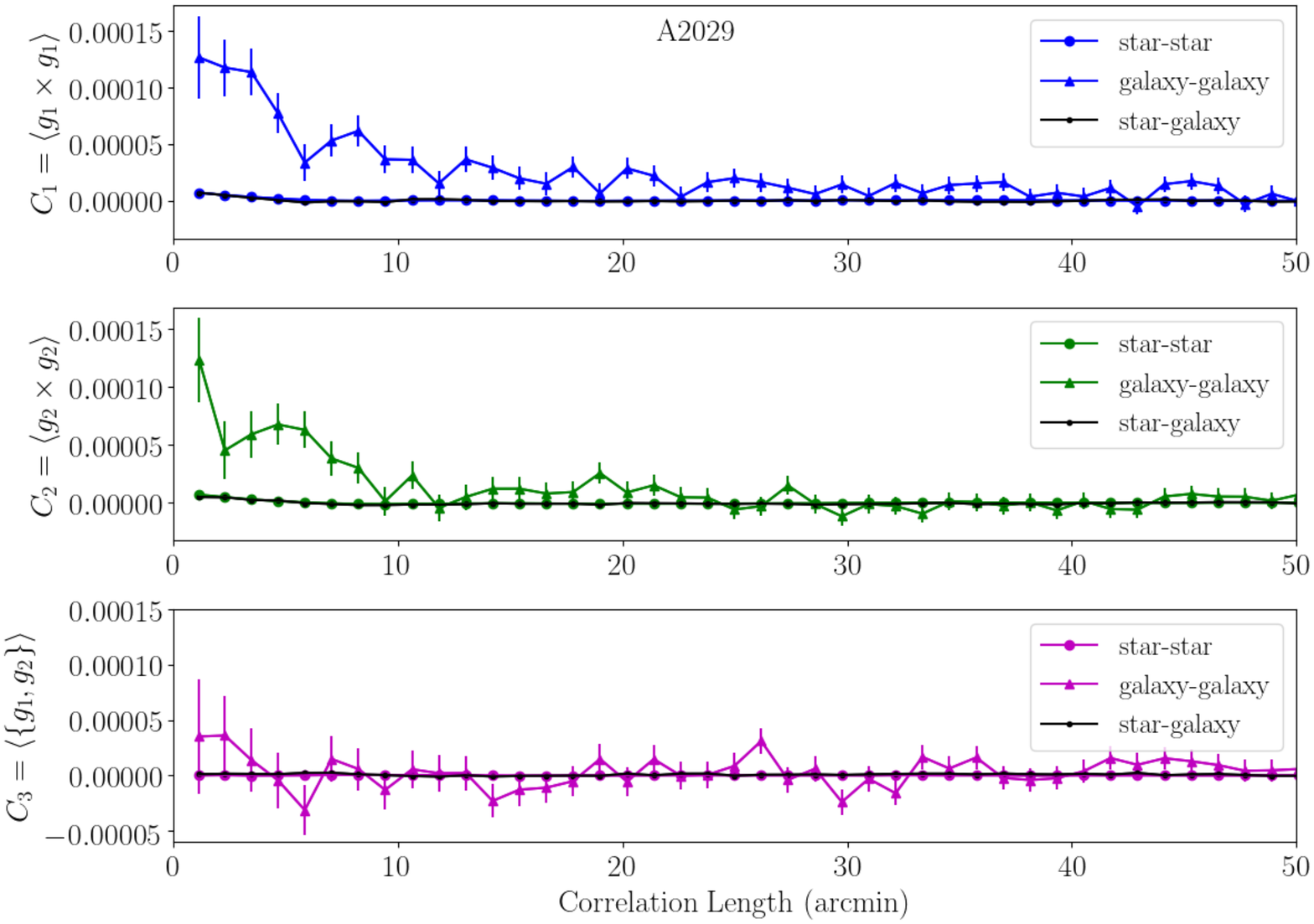}
\caption{{\small Same as Figure~\ref{fig:correlfuncs}, but for clusters A2457 (top) and A2029 (bottom).}}
\label{fig:correlfunc2}
\end{center}
\end{figure}

The quality of the PSF fits can be judged with two-point ellipticity correlation functions, given as
\begin{equation}
C_{i} = \langle e_i( {\bf r}) \times e_i( {\bf r+ \theta})\rangle,
\end{equation}
where $e_i$ is the {\it i}th ellipticity moment of an object at position r, and brackets denote an average over all pairs within a separation $\theta$. The $C_1$ and $C_2$ functions evaluated on galaxy pairs should have a relatively high amplitude, reflecting the imprint of cluster shear signal on galaxy shapes. In the limit of a successful PSF correction, the $C_1$ and $C_2$ functions should vanish when evaluated over star-star and star-galaxy pairs: the stars have been circularized and should have no ellipticity at all ($|\boldsymbol{e}|\sim0$), and galaxy ellipticites should not be correlated with rounded stars. The ``control'' cross-correlation function is given as  
\begin{equation}
C_{3} = \langle e_1( {\bf r}) \times e_2( {\bf r+
  \theta})+e_2( {\bf r}) \times e_1( {\bf r+ \theta})\rangle,
\end{equation}
and in the absence of systematic errors in the PSF should be consistent with zero over all pairs of objects.  The set of star-star, galaxy-galaxy and star-galaxy correlation functions are shown in Figure~\ref{fig:correlfuncs} for Abell 85 and Abell 1606, and in Figure~\ref{fig:correlfunc2} for Abell 2029 and Abell 2457. All correlation functions in Figures~\ref{fig:correlfuncs} and~\ref{fig:correlfunc2} show the anticipated behavior: the galaxy-galaxy autocorrelation functions dwarf the systematics probed by the star-star and star-galaxy correlations, and the amplitude of the ``test function'' C3 is ten times lower than C1 and C2. Accordingly, no systematics in PSF correction are apparent in these figures.  


\subsection{Lensing Aperture Mass Maps}\label{sec:ConvergenceMapping}
%, but also covers the smallest area, $2.53$ square degrees, due to a comparatively tighter dither pattern during observations. 
To extract cluster WL signal from the tangential ellipticities of background galaxies, we employ the software developed by \cite{huwethesis} and used in JM15 to produce shear maps of Abell 3128. 

A series of aperture mass maps are constructed with progressively larger Schirmer filter radii ($3000 \le R_S \le 9000$ pixels or 13\arcmin-42\arcmin). The cluster's aperture mass ($M_{\rm ap}$) signal is maximized at some $R_S$. For this paper, we deem ``significant" any $M_{\rm ap}$ peaks $\geq 4\sigma$ within 0.5 Mpc of the known X-ray center of the cluster, i.e. the cluster virial radius.

For computational efficiency, galaxy catalogs are first binned into 200 by 200 spatially adjacent blocks (the "pixels" in the final $M_{\rm ap}$ map), and an average reduced shear $g_1$ and $g_2$ for galaxies in that block is computed. Block by block, the aperture mass statistic of Eqs.~\ref{eqn:aperturemassstat} (with the Schirmer filter of~\ref{eqn:Schirmerfilter}) is computed to obtain a 2-D mass map of the cluster. As a test for systematic errors, B-mode maps are made by substituting Eq.~\ref{eqn:gc} into Eq.~\ref{eqn:aperturemassstat}. To obtain a signal-to-noise for both E- and B-mode $M_{ap}$ maps,  a random-noise aperture
mass map is generated by computing aperture mass statistic on
a catalog of shuffled galaxy positions, and taking the variance
of 100 such noise realizations.

Taking the variance of noise maps assumes a Gaussian distribution of pixel values. The assumption is weakly justified, however: the Schirmer filter kernel averages over nearly the same galaxy sample between adjacent blocks. The result is that adjacent pixels in the aperture mass maps are highly correlated with one another, and the S/N of any one pixel is ambiguous \citep{2003AJ....125.1014J}. Accordingly, we make significance or ``sigma'' $M_{\rm ap}$ maps, as follows. A very large number of noise maps is generated, and at every 200-pixel block of the observation, the number of noise maps with greater WL signal than the true $M_{\rm ap}$ signal map is counted. This number is converted into a Gaussian-type confidence $\sigma$ that quantifies the significance of the shear signal in that pixel block. The maximum attainable $\sigma$ depends on the number of noise iterations. In this study we generate roughly 1,000,000 random maps per Schirmer filter, corresponding to a maximum detection confidence of 4.8~$\sigma$.

\subsection{Parametric Mass Fitting}\label{sec:massnorm}

Aperture mass maps return only the relative mass enhancements in an observation, not the physical mass contained in the cluster. To obtain mass normalizations of the maps, axisymmetric NFW weak lensing shear profiles are fit to the galaxies' two-dimensional tangential ellipticity signal. The prescription of~\cite{2000ApJ...534...34W} is used to compute the halo’s reduced shear for a given $M_{200c}$ at the location of every background galaxy, with halo concentrations from \cite{2013ApJ...766...32B}. We find the clusters' best-fit $M_{200c}$ by minimizing $\chi^2$ between the NFW halo’s shear profile and galaxy $e_{\tan}$.
%A full description of the mass fitting procedure is given in JM15. 

All NFW shear profiles are centered on the highest $\sigma$ pixel of the clusters in the aperture mass sigma maps. Due to our binning scheme, each $M_{\rm ap}$ map pixel spans 200 pixels (53\arcsec) on the observation (cf. Section~\ref{sec:ConvergenceMapping}). The ambiguity in what is reported as center of a WL peak can bias mass estimates through mis-centering of the tangential ellipticity signal. More seriously, the true center of the weak lensing signal is ill-defined because the observed peak of the weak lensing signal is in reality the combination of cluster shear and galaxy shape noise. Centering NFW profiles on the highest $\sigma$ pixel will necessarily bias the resulting mass high, because a profile is being fit to where the shape noise has a positive tangential alignment.

We quantify this bias on the reported cluster masses with a set of simulations with mock shear catalogs that are based on the real galaxy position and shape noise distributions.  We begin with a real galaxy catalog (here, Abell 2029), and replace the observed galaxy ellipticities with a combination of the tangential shear from a cluster-sized NFW halo and a random ellipticity to mimic shape noise. %The model halo's NFW shears are computed at each catalog $(x,y)$ position with the prescription of~\cite{2000ApJ...534...34W}. 
Random ellipticities are drawn from a Gaussian with variance equal to the mean ellipticity of galaxies in the real catalog: $N(0,\langle \epsilon \rangle= 0.45)$. The NFW halo is fixed in mass, redshift and position on the “observation,” so the only change between mock catalog realizations is the shape noise assigned to each catalog entry.  

One thousand mock shear catalogs are created, and aperture mass maps are computed for each with a Schirmer filter radius of 10,000. We then record the distribution of offsets between the fiducial NFW centroid and the peak S/N pixel in each mock $M_{\rm ap}$ map; the variance is the uncertainty in the cluster’s true centroid due to shape noise. 

To turn this centroid uncertainty into an uncertainty on cluster mass from shape noise and address the question of bias, we take 4,000 random perturbations of the (now real) cluster centroid within a radius defined by the centroid uncertainty from simulations, and recompute the NFW mass. The median of the distribution of masses is the ``true,'' de-biased cluster mass, and the discrepancy between it and the mass obtained by naively centering on the peak $\sigma$ pixel in the real $M_{\rm ap}$ maps is a measure of the severity of the bias. 

Results of simulations for three different cluster masses, corresponding to the three different mass regimes in our cluster sample, are presented in Section~\ref{subsec:nfwtests}.

\section{Results}\label{sec:Results}

\subsection{Identification of High Significance $M_{\rm ap}$ Peaks}\label{subsec:MassMaps}
\begin{deluxetable}{ccccc}
\tabletypesize{\scriptsize}
\tablewidth{0pt} 
\tablecaption{List of Cluster Detections\label{tab:clustersignal}}
\tablehead{
\colhead{Cluster} & \colhead{$\alpha$}& \colhead{$\delta$}&\colhead{S/N}&\colhead{Detection Significance} \\
\colhead{} & \colhead{(J2000.0)} & \colhead{(J2000.0)}&\colhead{} &\colhead{}\\
} 
\startdata
{A85}& 0$^{h}$41$^{m}$48$\fs$4 & -9$\degr$18$\arcmin$16$\arcsec$ & 5.8 & $\ge$ 4.80 $\sigma$ \\
\hline
{A2029}& 15$^{h}$10$^{m}$56$\fs$4& +5$\degr$44$\arcmin$58$\arcsec$ & 5.8 & $\ge$ 4.89 $\sigma$\\
\hline
{A1606}& 12$^{h}$44$^{m}$45$\fs$1 & -11$\degr$44$\arcmin$03$\arcsec$ & 5.5 & $\ge$ 4.89 $\sigma$\\ 
\hline
{A2457}& 22$^{h}$36$^{m}$48$\fs$6& +1$\degr$37$\arcmin$56$\arcsec$ &4.1 & 3.75 $\sigma$\\ 
\enddata
\tablecomments{\scriptsize{Centers of the cluster weak lensing signals.}}
\end{deluxetable}

We report the weak lensing signal of all four clusters, with $M_{\rm ap}$ significance maps presented here. Cluster mass maps are shown with the Schirmer filter size that maximizes the detection significance. For reference, all maps are plotted wtih a 10\arcmin~scale bar, which spans a physical scale between 650 kpc and 1.3 Mpc depending on the distance to the cluster. We also compare our cluster WL signal with X-ray gas and optically-identified knots of galaxies. The WL centroids reported in Table~\ref{tab:clustersignal} are the location of the pixel with the highest $\sigma$. 

\begin{figure}[htbp]
\begin{center}
\includegraphics[width=0.7\textwidth]{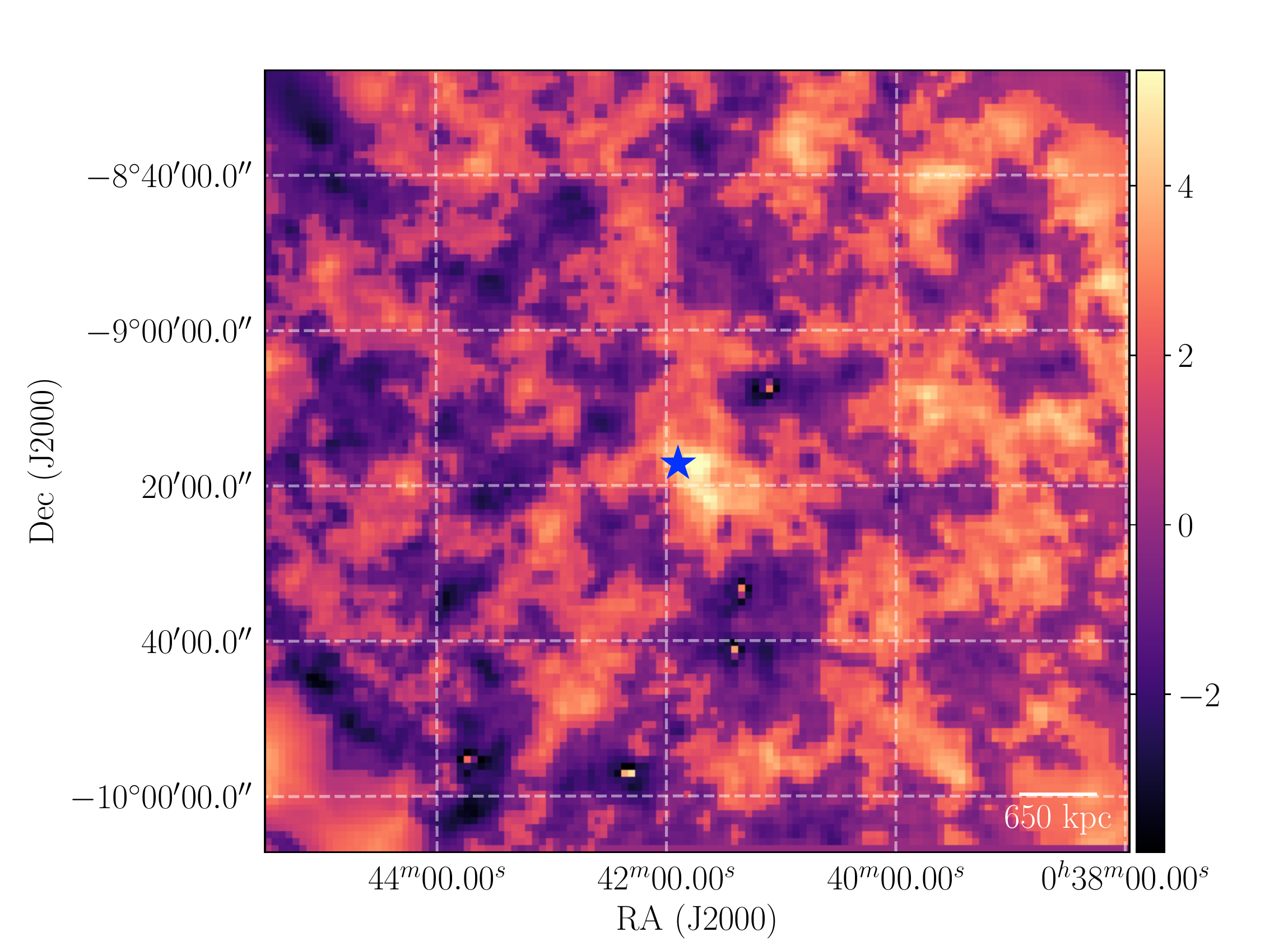}
\caption{{\small A85 significance maps made with Schirmer filter sizes of 4000 pixels. The color scale represents the significance of detection, and the star marks the position of the BCG.}}
\label{fig:A85sigma}
\end{center}
\end{figure}

Aperture mass sigma maps for Abell 85 are shown in Figure~\ref{fig:A85sigma}. The figure shows the significance of an $M_{ap}$ map with Schirmer filter size of 4000 pixels, which maximizes the cluster's lensing signal. The A85 WL signal is detected with $\sigma=4.80$ and $S/N = 5.8$. The lensing signal has a northeast-southwest alignment, which is seen in X-ray studies such as~\cite{2002ApJ...579..236K}, \cite{2005A&A...432..809D}, and \cite{2015MNRAS.448.2971I}.

\begin{figure}[tb!]
\begin{center}
\includegraphics[width=0.75 \textwidth]{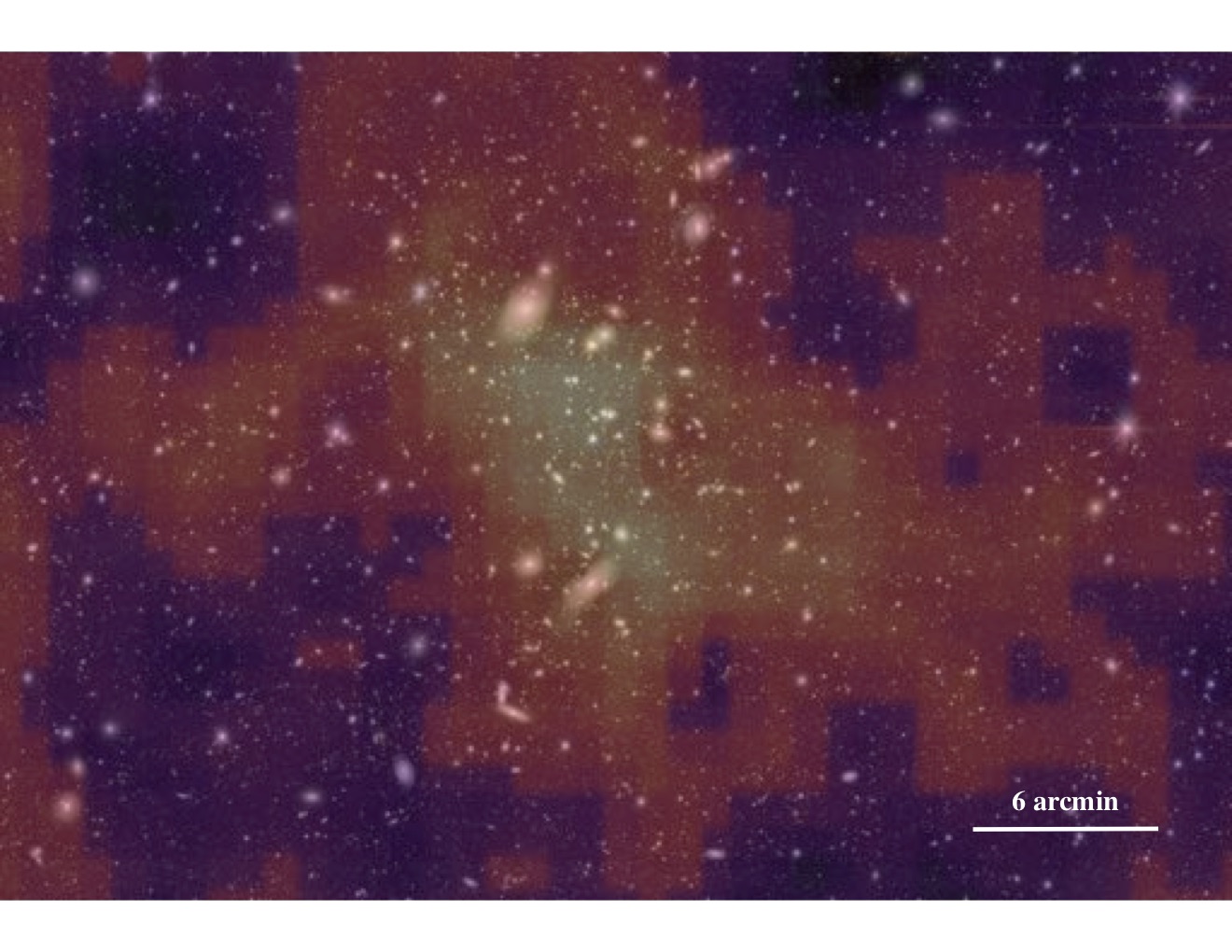}
\caption{\small{A85 $R_{\rm S}$=4000 significance map superimposed on $gri$ composite.}}
\label{fig:A85overlay}
\end{center}
\end{figure}

Figure~\ref{fig:A85overlay} shows that the BCG of the cluster (0$^{h}$41$^{m}$34\fs$9, -9$\degr$21$\arcmin$50\arcsec$ and marked with a blue star) is 3$\arcmin$~away from the center of the WL signal. The WL signal peak itself closely coincides with the published X-ray center of $0^{h}41^{m}50\fs1$, $-9\degr$18$\arcmin$36$\arcsec$. Given the uncertainty in the WL centroid for an Abell 85-mass cluster ($\sim 1\farcm1$, see Section~\ref{subsec:nfwtests}), this offset is potentially significant.

\begin{figure}[htb]
\begin{center}
\includegraphics[width=0.75\textwidth]{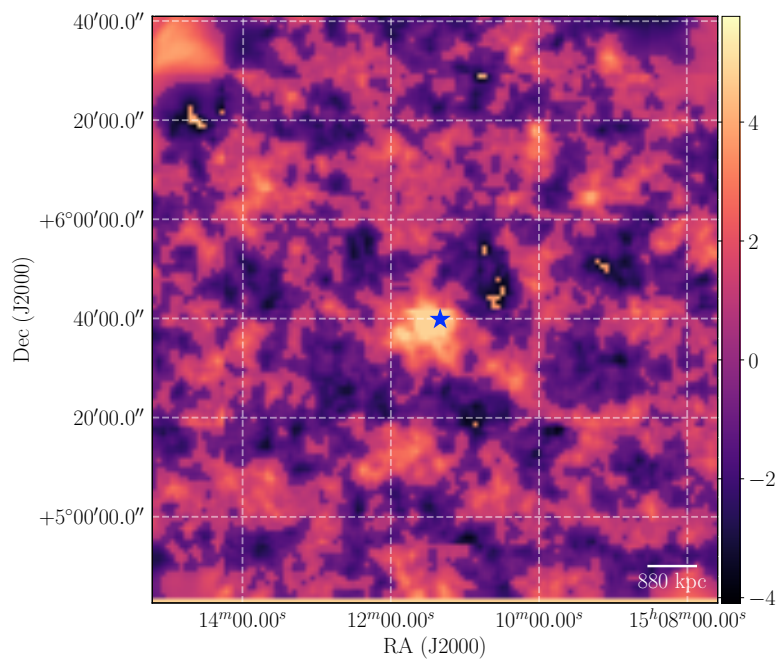}
\caption{\small{A2029 significance maps made with Schirmer filter sizes of 4000 pixels. The color scale indicates the significance of detection, and the star marks the location of the BCG.}}
\label{fig:A2029sigma}
\end{center}
\end{figure}

Significance maps for Abell 2029 are displayed in Figure~\ref{fig:A2029sigma}. At $R_S = 4000$ pixels and higher, A2029 saturates our detection significance of 4.89$\sigma$, and has E-mode $S/N \sim 5.8$. The cluster weak lensing signal covers an area of 15\arcmin, or about 1.4 Mpc at the redshift of A2029.

\begin{figure}[htb]
\begin{center}
\includegraphics[width=0.75\textwidth]{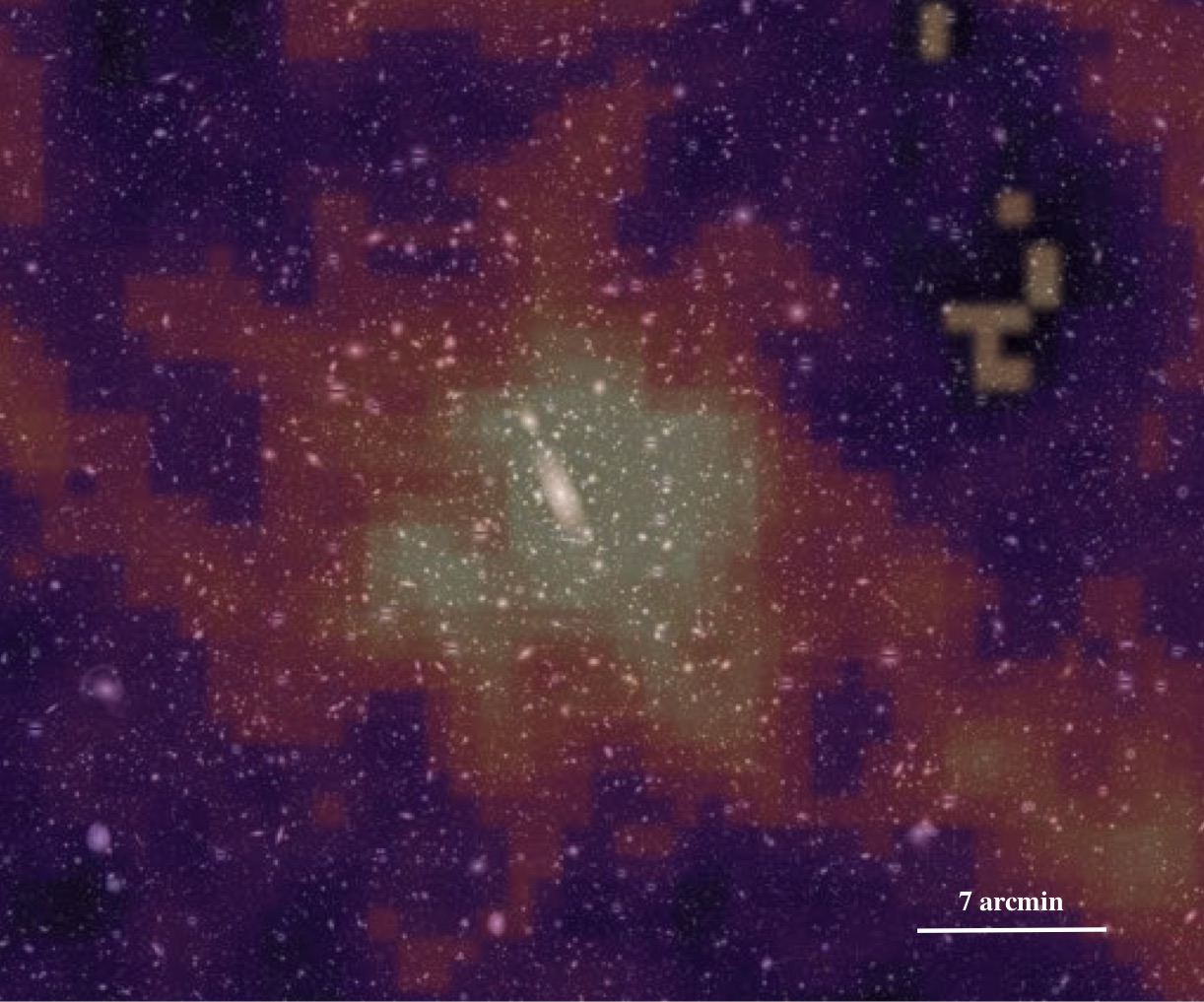}
\caption{Abell 2029 $R_{\rm S}$=4000 pixel significance map superimposed on $gri$ composite. The white scale bar spans about 3\arcmin~across the observation.}
\label{fig:A2029overlay}
\end{center}
\end{figure}

Figure~\ref{fig:A2029overlay} shows the Abell 2029 $R_{\rm S}$=4000 pixel significance map overlaid on a $gri$ composite image. The center of the A2029 weak lensing signal is clearly aligned with the BCG of A2029, and also encompasses several other galaxies at the cluster redshift. X-ray studies of A2029~\citep{2012MNRAS.422.3503W, 2013ApJ...773..114P} confirm the smooth distribution and size scale of the observed A2029 that we observe, as well as a roughly NE-SW orientation.

\begin{figure}[htbp]
\begin{center}
\includegraphics[width=0.75\textwidth]{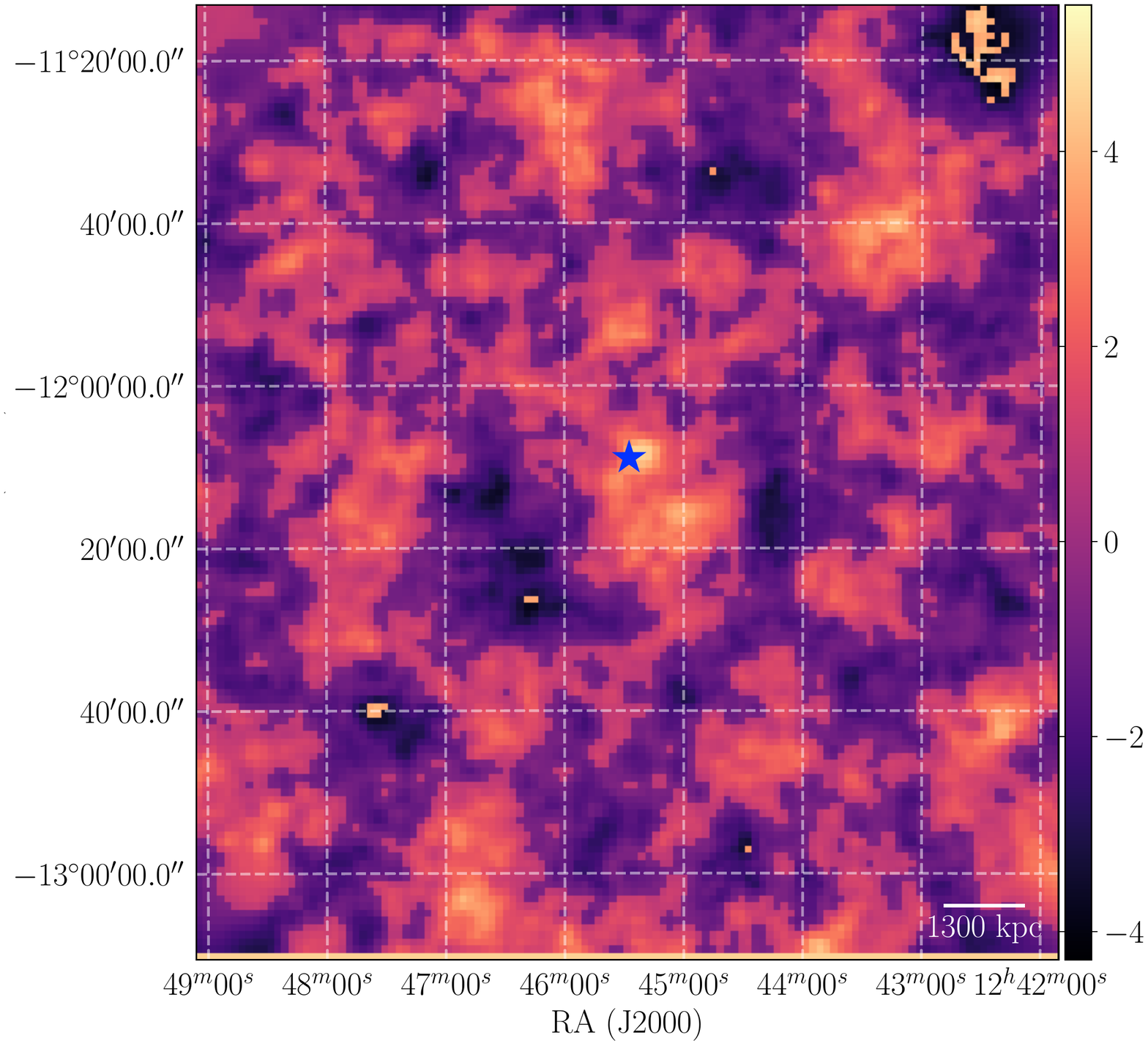}
\caption{\small{A1606 significance maps made with $R_{\rm S}=7500$ filter. The color represents the significance of detection, and the blue star marks the location of the BCG.}}
\label{fig:A1606sigma}
\end{center}
\end{figure}

\begin{figure}[htb]
\begin{center}
\includegraphics[width=0.75\textwidth]{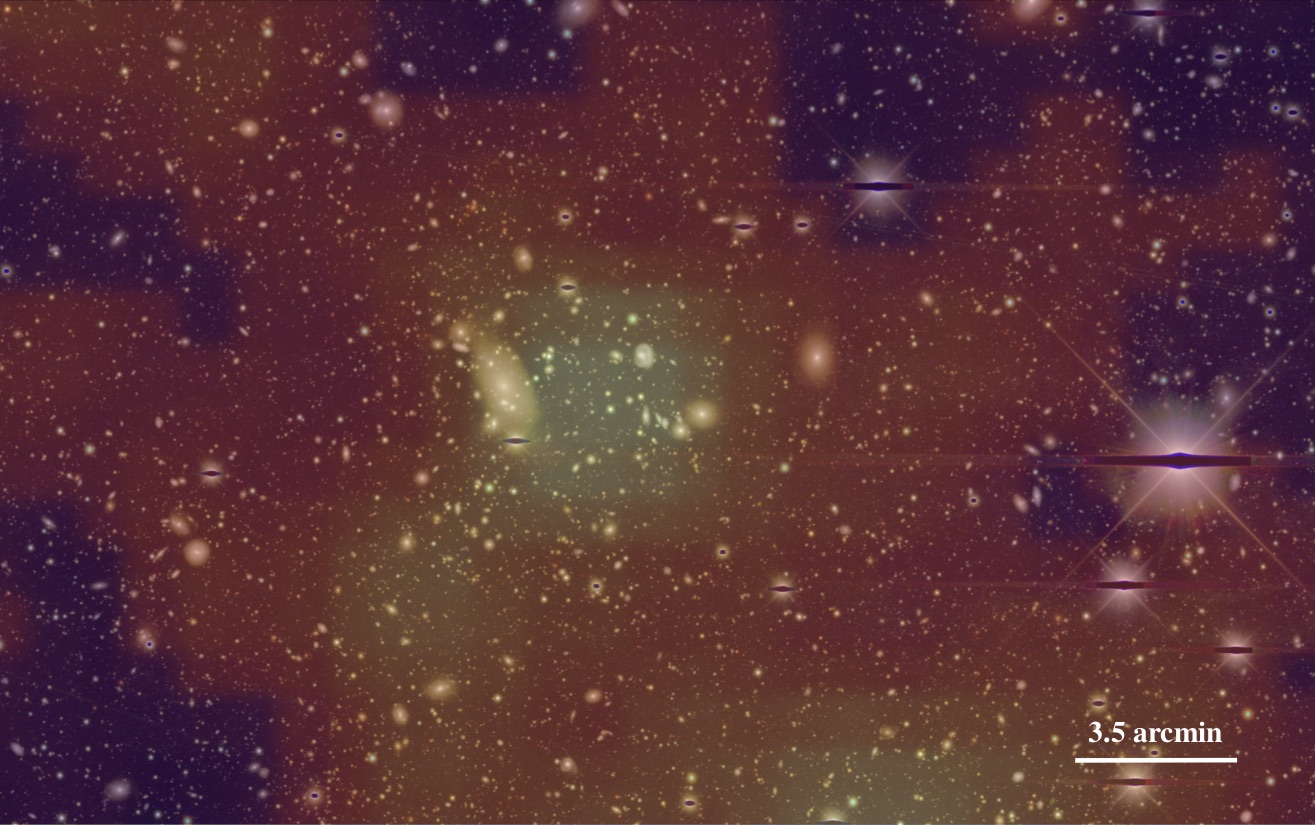}
\caption{\small{Abell 1606 $R_{\rm S}$=6000 pixel significance map superimposed on $gri$ composite image.}}
\label{fig:A1606overlay}
\end{center}
\end{figure}

The weak lensing signal of Abell 1606 is distinctive for its central concentration, as shown in Figure~\ref{fig:A1606sigma}, and the signal encompasses the brightest cluster galaxy (Figure~\ref{fig:A1606overlay}). It saturates our significance maps with 4.89~$\sigma$ and attains its maximum S/N of 5.5 in the $R_{\rm S}$=7000 $M_{\rm ap}$ maps (Figure~\ref{fig:A1606sigma}). 
 
\begin{figure}[htbp]
\begin{center}
\includegraphics[width=0.75\textwidth]{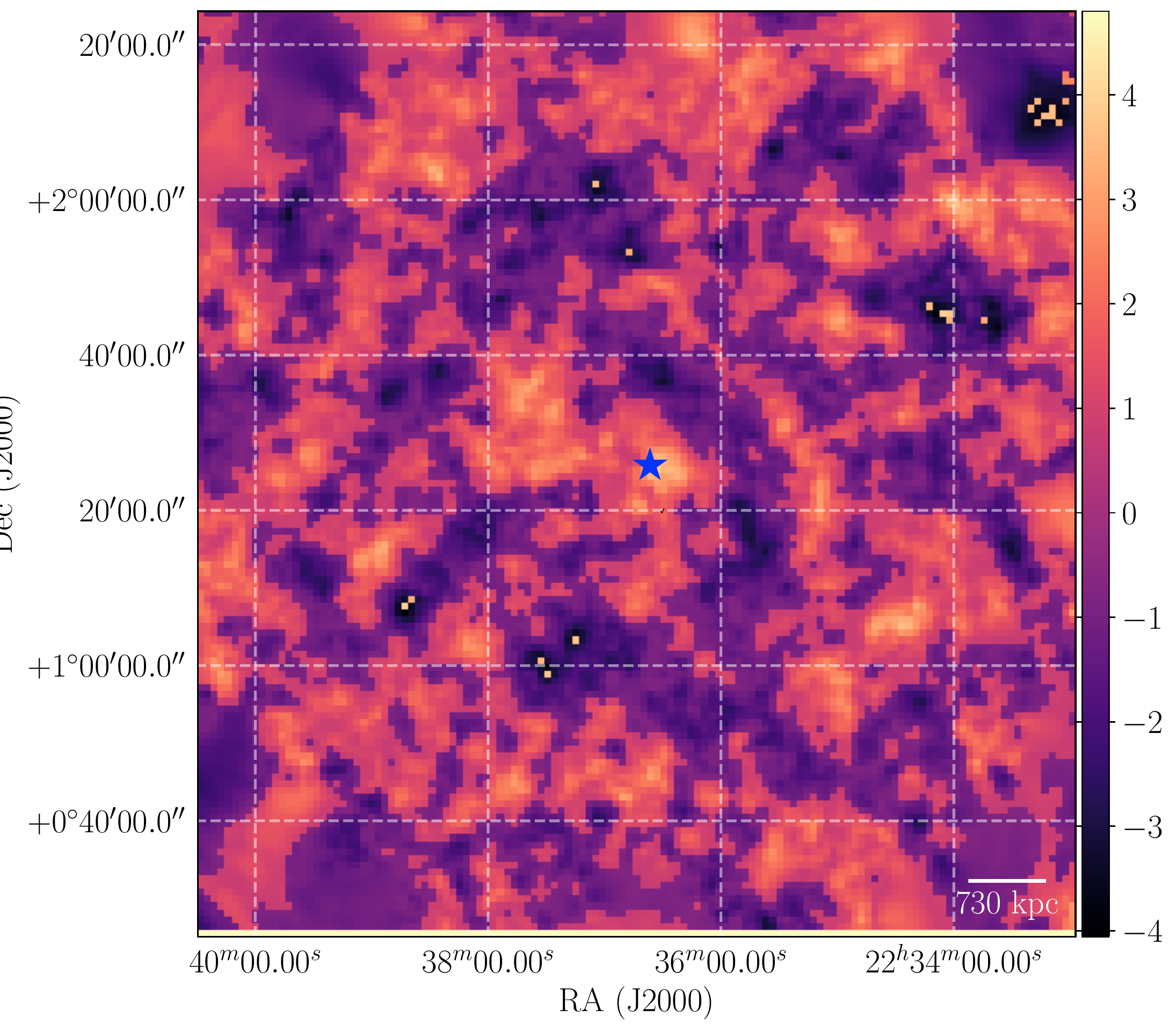}
\caption{\small{Significance maps of A2457 made with Schirmer filter radii of 3500 pixels. The plot color represents the significance of detection, and the scale bar spans 3\arcmin~on the original observation.}}
\label{fig:A2457sigma}
\end{center}
\end{figure}

\begin{figure}[htb]
\begin{center}
\includegraphics[width=0.75\textwidth]{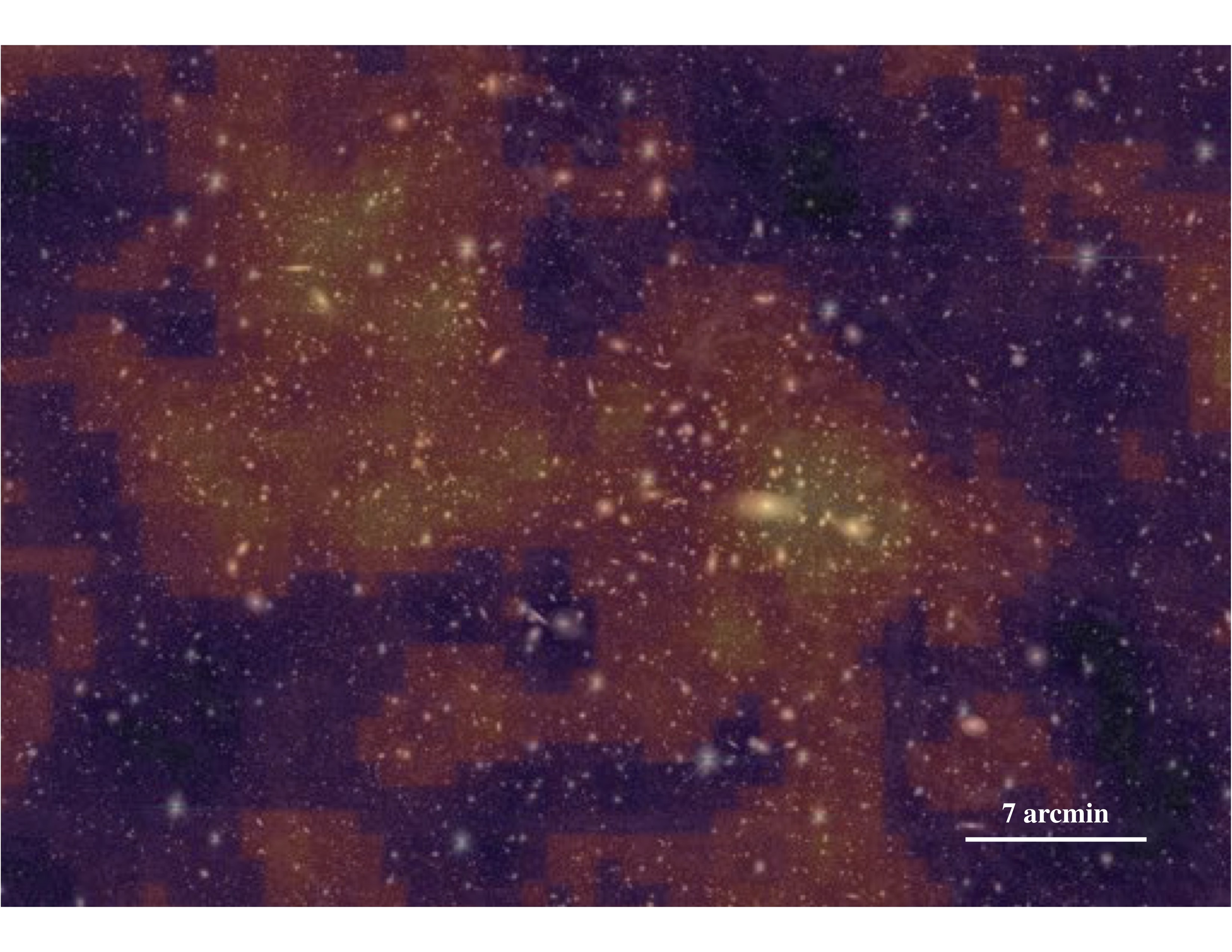}
\caption{\small{Abell 2457 $R_{\rm S}$=3500 pixel significance map superimposed on $gri$ composite.}}
\label{fig:A2457overlay}
\end{center}
\end{figure}

Figure \ref{fig:A2457sigma} shows significance maps for Abell 2457. The cluster reaches its maximum significance of $\sigma=3.75$ at $R_{\rm S}$=3500 pixels, which corresponds to $1.8\arcmin$ on the observation. In corresponding E-mode ($g_{\tan}$) S/N maps, the cluster is detected at S/N=4.1. Both the maximum S/N and the maximum detection significance of $\sigma=3.75$ are lower than the other 3 clusters. This is likely attributable to the relatively smaller mass of the cluster, cf. Table~\ref{tab:NFW1}. 
Peak significance appears to be aligned with the cluster BCG, but the rest of the signal has a noticeable east-west alignment, consistent in reconstructions across all Schirmer filter scales. The east-west configuration of A2457's WL signal is supported by the arrangement of galaxies visible in Figure~\ref{fig:A2457overlay}. X-ray studies of A2457 tend to concentrate their efforts near the BCG, but also report an east-west elongation of the X-ray gas \citep{2014AJ....147..156L}. 

\subsection{NFW Shear Profile Fits}\label{sec:nfwfits}
\begin{deluxetable*}{ccccc}
\tabletypesize{\scriptsize}
\tablewidth{0pt} 
\tablecaption{Masses from NFW Profile Fits  \label{tab:NFW1}}
\tablehead{
\colhead{Cluster} & \colhead{$\alpha$}& \colhead{$\delta$} & \colhead{$M_{200c}^{a}$} & \colhead{$\Delta M_{\rm cent}^{b}$}\\
\colhead{} & \colhead{(J2000.0)} & \colhead{(J2000.0)} & \colhead{$10^{14} M_\odot$}& \colhead{$10^{14} M_\odot$}
} 
\startdata
{A85}&  0$^{h}$41$^{m}$45$\fs$4 & -9$\degr$20$\arcmin$31$\arcsec$ & $3.63^{+1.24}_{-0.91}$ & $3.23^{+0.37}_{-0.55}$\\
{A2029}&  15$^{h}$11$^{m}$02$\fs$0& +5$\degr$43$\arcmin$33$\arcsec$ & $12.2_{-1.8}^{+1.6}$& $11.9^{+0.30}_{-0.50}$\\
{A1606}& 12$^{h}$44$^{m}$34$\fs$0 & -11$\degr$59$\arcmin$59$\arcsec$& $4.43_{-1.26}^{1.36}$& $4.06^{+0.37}_{-0.55}$\\
{A2457}& 22$^{h}$35$^{m}$31$\fs$5 & +1$\degr$36$\arcmin$17$\arcsec$ & $1.70_{-0.656}^{+0.872}$& $\lesssim 0.82$\\ 
\enddata
\tablenotetext{a}{Equivalent one-sigma statistical uncertainties}
\tablenotetext{b}{Mass from centroid shuffling procedure (Section~\ref{sec:massnorm}) with equivalent one-sigma systematic uncertainties. }
%\tablecomments{\scriptsize{The coordinates listed above are the centroids chosen for the NFW shear profile fits. Equivalent one-sigma uncertainties are shown as superscripts and subscripts. }}
\end{deluxetable*}

Following the procedure in Section~\ref{sec:ConvergenceMapping}, we find mass normalizations for the cluster significance maps presented above. NFW shear profiles are centered at the highest-$\sigma$ pixel of the aperture mass maps and fit to the entire background galaxy catalog. Resulting masses presented in the fourth column of Table~\ref{tab:NFW1}. To obtain an uncertainty, we take 1,000 bootstrap resamples of 50\% of the total background galaxy catalog and sum 34.1\% of the returned masses on either side of the distribution to obtain equivalent 68\% confidence interval. The errors are adjusted by a factor of $1/\sqrt{2}$ to account for the 50\% resampling. The equivalent fractional uncertainty on the masses is 13\% for the most massive cluster Abell 2029, around 30\% for Abell 1606 and Abell 85, and up to 52\% on the high end for Abell 2457, the lowest mass cluster in the sample. %On the low end, the fractional uncertainty of Abell 2457 is 35\%, but this should be interpreted as a cluster not being able to have less than zero mass, i.e. the mass is within 1$\sigma$ of our detection limit. 

\begin{figure}[htb]
\begin{center}
\includegraphics[width=\textwidth]{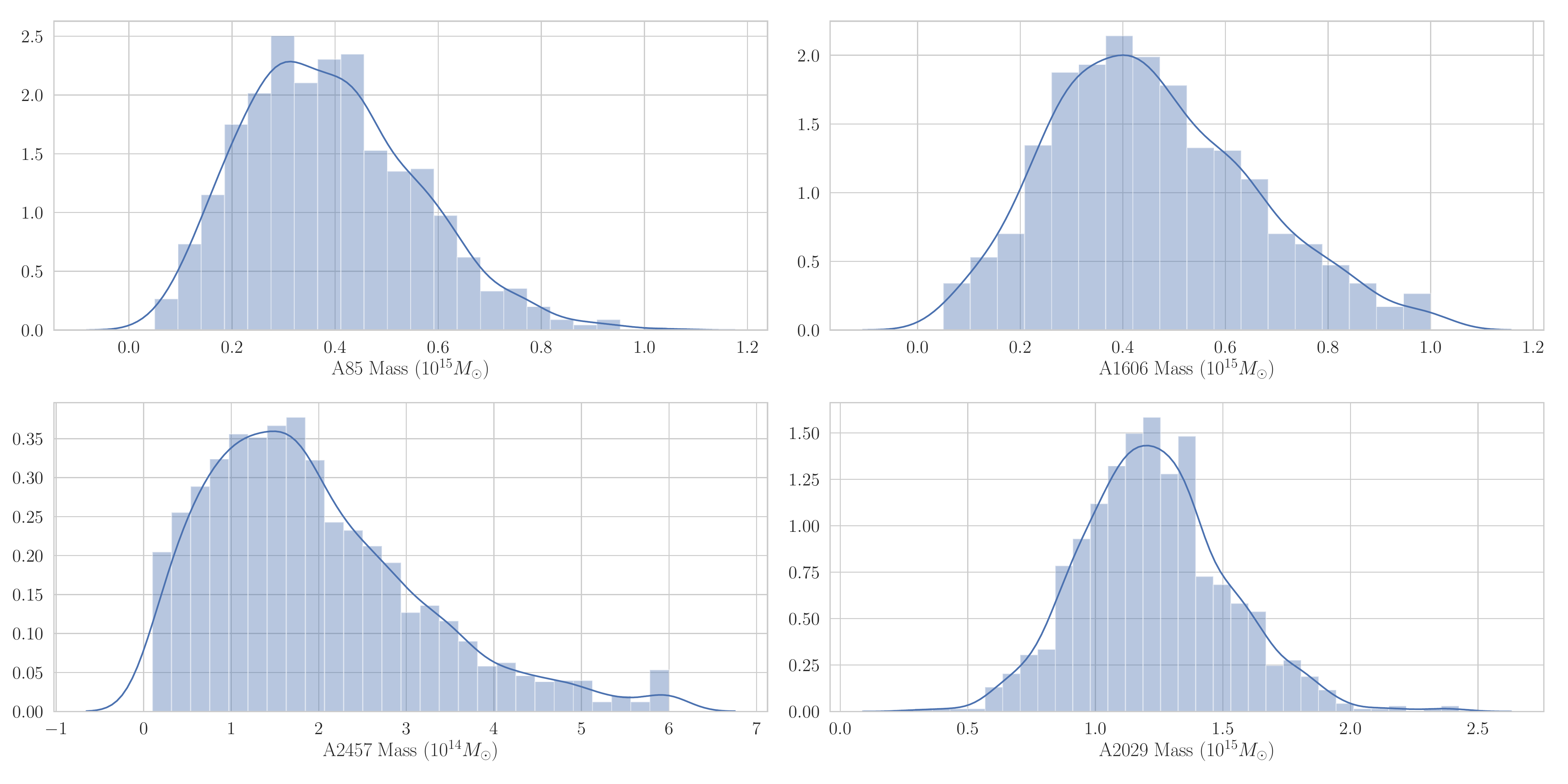}
\caption{\small{Distribution of best-fit masses returned by the bootstrap resampling of the background galaxies of all four clusters. Solid lines are kernel density estimates of the PDF. Units on the y-axis are arbitrary, but represent frequency.}}
\label{fig:masshist}
\end{center}
\end{figure}

The output of the mass resampling is shown in Figure~\ref{fig:masshist} with the kernel density estimates from the Seaborn data visualization package. %Kernel density estimates (KDE) are non-parametric models of the probability density function of a random variable; in the Gaussian KDEs used here, each data point contributes a Gaussian curve to the total. 

\subsection{Systematics Tests \label{subsec:nfwtests}}
\begin{figure}[htb]
\begin{center}
\includegraphics[width=0.75\textwidth]{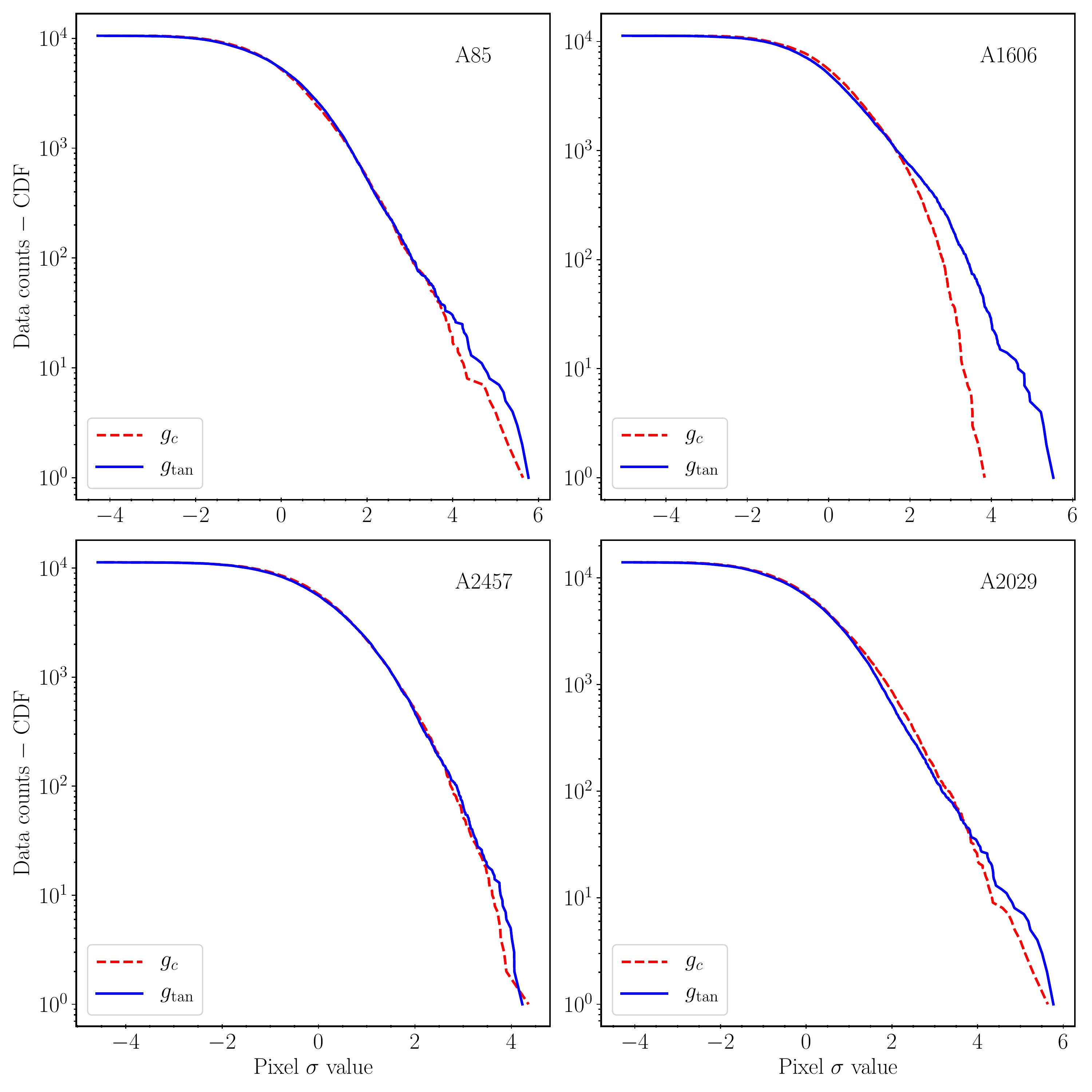}
\caption{\small{``Survival function'' of E-mode ($g_{\tan}$) and and B-mode ($g_c$) $\sigma$ pixel values for observed clusters.}}
\label{fig:cdf}
\end{center}
\end{figure}

%In Section~\ref{sec:massnorm} we discuss the systematic errors to which parametric mass fits are subject. We investigate the possible effects of several of these systematics on our results in this section. 
Our analysis depends on the detection of $\sim 4 \sigma$ peaks in aperture mass maps, but the look-elsewhere effect guarantees that some high-significance peaks will appear regardless of the presence of any shear signal. Moreover, the smaller the $R_S$ used in the $M_{\rm ap}$ maps, the more ``samples'' are taken and the more likely high-significance peaks are to emerge. Figure~\ref{fig:cdf} illustrates a simple test for this effect. As discussed in Sections~\ref{sec:theory}, PSF systematics are expected to add equal power to tangential ($g_{\tan}$ or E-mode) and cross shear ($g_c$ or B-mode) lensing signal. If the lensing signal is genuine, the distribution of tangential shear map (E-mode) pixel values should have an excess of high S/N pixels relative to the cross-shear B-mode maps. A so-called ``survival function," or difference of data counts and the cumulative distribution function as a function of data values, is shown in Figure~\ref{fig:cdf}. The area underneath E-mode curve is 4.3\% higher than B-modes for A2029; 5.8\% for A85; 15.9\% higher for A1606; and 2.2\% for A2457. Importantly, the excess power occurs at the high-end ($S/N>3$) tail of the distribution, which indicates that look-elsewhere effects or other systematics are less important than cluster signal.

\begin{figure}[thb]
\begin{center}
\includegraphics[width=0.9\textwidth]{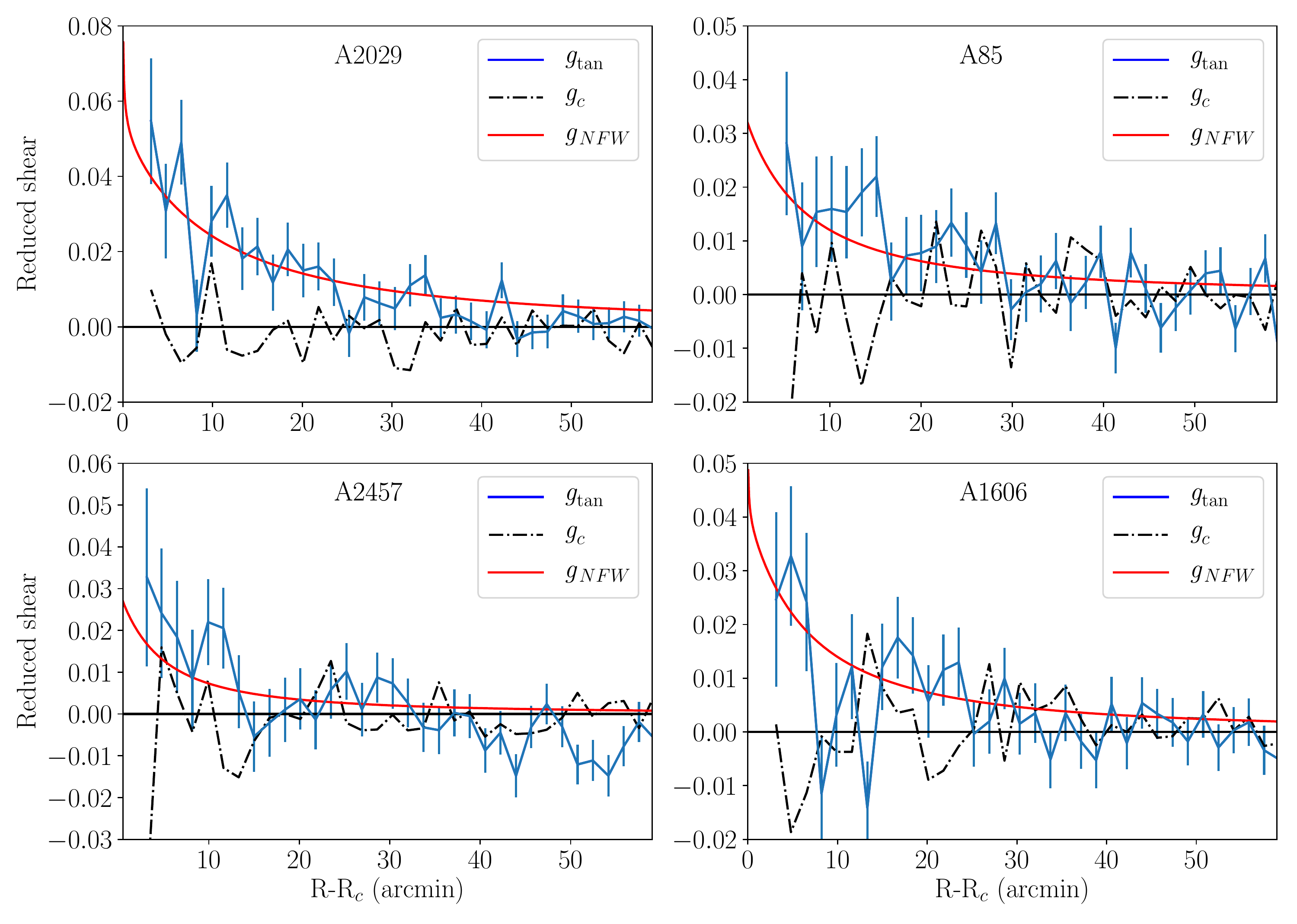}
\caption{\small{Tangential shear profiles of NFW masses in Table~\ref{tab:NFW1} (red line), overplotted on the azimuthually averaged tangential shear ($g_{\tan}$) of background galaxies (solid blue line). %The error bars on tangential ellipticity are the value of reduced shear divided by $\sqrt{N_{\rm gal}}$ in each annulus. Any contribution by systematics to the WL signal of background galaxies is revealed by their 
Cross shear ($g_c$) signal is plotted as a dashed black line. Note that y-axis ranges differ between panels. $R-R_c$ is the distance from the observed WL centroid.}}
\label{fig:nfwAll}
\end{center}
\end{figure}
Figure~\ref{fig:nfwAll} offers another way to verify that the B-mode cross shear is consistent with zero. The figure shows best-fit NFW shear profiles from Section~\ref{sec:nfwfits} plotted against the azimuthally-averaged tangential ellipiticity of background galaxies for all clusters. %The wide area of the DECam observations and large number of background galaxies allow for a fine radial binning and detailed inspection of galaxy ellipticity signal. 
The ellipticity $g_{\tan}$ should peak at the cluster center ($R-R_c$=0) and as distance from the cluster center increases, the galaxy ellipticity signal should approach zero. In the absence of PSF residuals, the cross shear (B-mode) $g_c$ should be consistent with zero at all radii. Figure~\ref{fig:nfwAll} shows the expected behavior: the projected NFW fits agree well with the tangential ellipticity, which asymptotically approaches zero. Except at the smallest distances from the cluster center, where the small number of galaxies causes shape noise to dominate, the B-mode statistic $g_c$ is consistent with zero. 
\begin{figure}[tbh]
\begin{center}
\includegraphics [width=\textwidth]{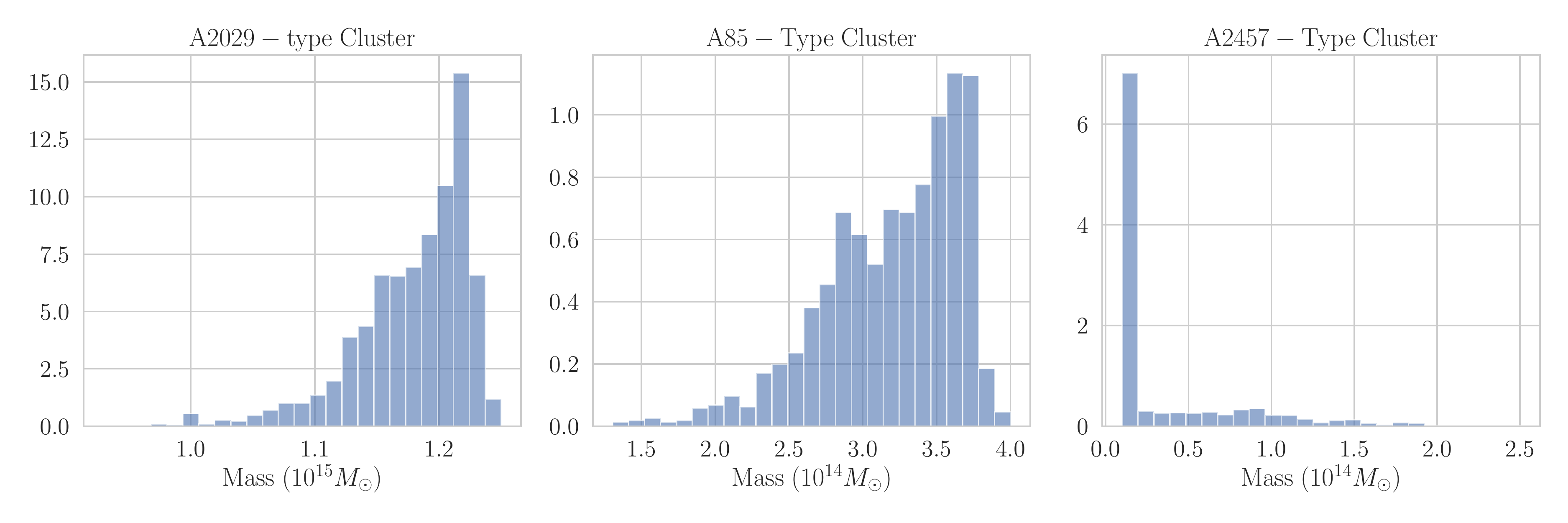}
\caption{\small{Distributions of best-fit NFW masses returned by the shape noise/centroid shuffling procedure (cf.~Section\ref{sec:massnorm})}}
\label{fig:randoffset}
\end{center}
\end{figure}
We implement the shape noise simulations of Section~\ref{sec:massnorm} in 3 representative regimes: an Abell 2029-type high mass cluster with $M=1.2\times10^{15}M_\odot$ at $z=0.077$; an Abell 85/Abell 1606-type intermediate mass cluster with $M=4\times10^{14}M_\odot$ at $z=0.06$; and an an Abell 2457-type low mass cluster with $M=1.6\times10^{14}M_\odot$ at $z=0.059$. Distributions of masses from the random perturbations of the real cluster centroids are shown in Figure~\ref{fig:randoffset}, and the resulting masses with equivalent $1\ \sigma$ uncertainties are shown in the fourth column of Table~\ref{tab:NFW1} as $\Delta M_{\rm cent}$.

The variance from shape noise in the mock high-mass cluster's centroid is only 124 pixels on the camera, or 32$\arcsec$. We obtain an equivalent de-biased Abell 2029 mass by randomly perturbing the fiducial center in Table~\ref{tab:NFW1} within a Gaussian distribution of $\sigma = 32\arcsec$, and re-computing the best-fit $M_{200c}$ to the (real) Abell 2029 catalog. The resulting mass is $11.9^{0.32}_{-0.57}\times 10^{14}M_\odot$. The difference with the fiducial Abell 2029 mass is only 2\%, 5 times smaller than the statistical uncertainty reported in Table~\ref{tab:NFW1}. 

In the intermediate-mass cluster simulation, the variance in the WL centroids is 253 pixels on the camera, or 66$\arcsec$. This variance is used to run 4,000 randomly shuffled NFW fits to the real A85 catalog, resulting in a de-biased A85 mass of $3.23^{0.37}_{-0.55}\times 10^{14}M_\odot$, 10\% lower than the fiducial A85 mass. The same procedure applied to Abell 1606 yields $M=4.06^{+0.37}_{-0.55}M_{\odot}$, 8\% lower than the fiducial mass. Here as well, the bias introduced by centering on the highest $sigma$ $M_{\rm ap}$ pixel is a fraction of the statistical uncertainties on the masses. 

The low-mass cluster simulation returned a significantly larger variance in the WL centroid: 2,760 pixels on the camera, or 12$\arcmin$. This is of order the size of the virial radius of Abell 2457 on the observation. Consequently, the random-perturbation mass fits to the Abell 2457 catalog returns only an upper bound: $M=\lesssim 0.818\times10^{14}M_\odot$. While this debiased mass is technically within one sigma of the fiducial A2457 mass in Table~\ref{tab:NFW1}, centering the NFW profile fit on the highest-$\sigma$ $M_{\rm ap}$ pixel clearly introduces a bias that is at least 50\% the mass of the cluster.

%Table~\ref{tab:centroidtests} shows the results of 1000 such shifts. 
%\begin{deluxetable}{cccc}
%\tabletypesize{\scriptsize}
%\tablewidth{0pt} 
%\tablecaption{Masses from Random Centroid Offsets %\label{tab:centroidtests}}
%\tablehead{
%\colhead{Cluster} & \colhead{$M_{200c}$} \\
%\colhead{} &\colhead{$10^{14} M_\odot$}
%} 
%\startdata
%{A85}&  $2.47\pm0.88$\\
%{A2029}&  $8.79\pm1.23$\\
%{A1606}& $2.77\pm0.62$\\
%{A2457}&  $1.07\pm0.24$\\ 
%\enddata
%\tablecomments{\scriptsize{Error bars are one standard deviation from the %mean.}}
%\end{deluxetable}

\subsection{Comparison to X-ray Masses\label{subsec:xraysnr}}
\begin{figure}[tbh]
\begin{center}
\includegraphics [width=0.6\textwidth]{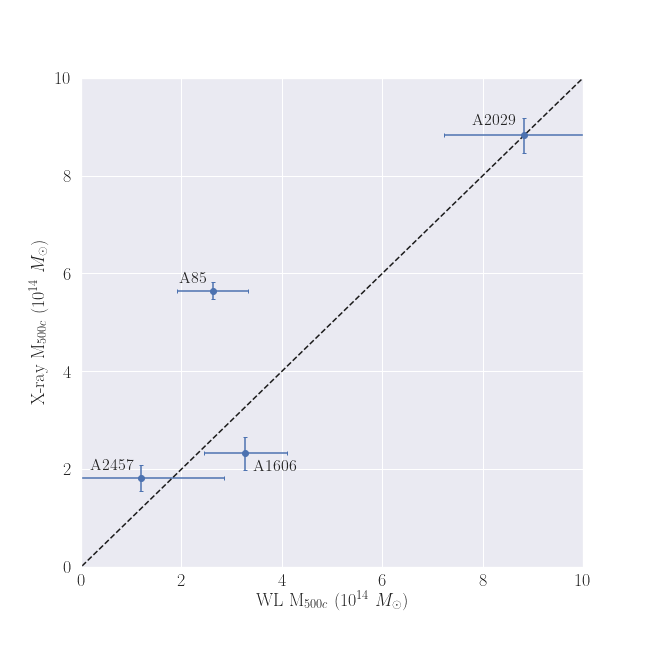}
\caption{\small{X-ray $M_{500c}$ from MXMC and \cite{2005A&A...432..809D} compared with our values of $M_{500c}$ for the clusters (converted from $M_{200c}$). The dashed line represents equal WL and X-ray masses.}}
\label{fig:xray}
\end{center}
\end{figure}

All the clusters considered in this work have been well-studied in X-rays; this was in fact a requirement in the target selection. As a consequence, all galaxy clusters in this paper have independent mass estimates, for which we queried the MCXC meta-catalog of X-ray cluster studies~\citep{2011A&A...534A.109P}. Abell 85 was the subject of a detailed X-ray study by~\cite{2005A&A...432..809D} in which they report a dynamical mass based on X-ray temperature. Because of the limited field-of-view of most X-ray telescopes, all X-ray quantities are measured out to $R_{500c}$. This radius defines the size within which the mean over-density of the cluster is 500 times the critical density at the cluster redshift. Mass estimates are thus based on the total matter contained within a sphere of radius $R_{500c}$ and assume hydrostatic equilibrium. To convert our $M_{200c}$-based masses into an equivalent $M_{500c}$, we used the conversions of~\citep{2003ApJ...584..702H}. Equivalent $M_{500c}$ WL masses from Table~\ref{tab:NFW1} and X-ray masses are plotted against one another in Figure~\ref{fig:xray}; the dashed line shows equal WL and X-ray masses. While conclusive statements cannot be made with a sample of four clusters, the weak lensing and X-ray masses appear consistent.

\section{Discussion and Conclusions}\label{sec:Discussion}

We believe that we are successful in observing the weak lensing signal of these four low-redshift clusters, and fitting their lensing signal with parametric masses. In our pilot study (JM15), we began with one of the largest clusters in the local Universe, A3128. In this study, we examine a similarly massive cluster (A2029), but also the lower-mass clusters A1606, A85 and A2457. In particular, the weak lensing maps and masses for Abell 1606 and Abell 2457 are the first in the literature. That said, the Abell 2457 detection is marginal: it is only a $\sigma = 3.75$ detection, with a statistical uncertainty of 50\% on the NFW mass. Moreover, the shape noise simulations of Section~\ref{subsec:nfwtests} show that the location of the centroid of a $\sim 1-2\times10^{14} M_{\odot}$ cluster is dominated by shape noise fluctuations in our observations. The large uncertainty in the position of the halo center means that centering the profile on the highest sigma $M_{\rm ap} pixel$ introduces a large bias on the Abell 2457 mass, and so the mass constraint on the cluster is not significant. With these caveats in mind, we do detect some weak lensing signal in the vicinity of the BCG and X-ray emission, and the NFW mass we recover for Abell 2457 is in agreement with the X-ray mass~(Figure \ref{fig:xray}. These factors suggest that the Abell 2457 detection is an upper limit, and that future observations should target higher mass clusters. 

Though Figure~\ref{fig:xray} shows generally good agreement between X-ray and WL masses, any of the following could contribute to the scatter: departures from hydrostatic equilibrium in X-ray modeling due to e.g. non-thermal pressure support~\citep{2008MNRAS.384.1567M}; halo triaxiality/departures from NFW spherical symmetry~\citep{2019MNRAS.490.4889H}; and photometric redshift uncertainties~\citep{2008ApJ...689..709O}. Expanding our sample to more low-redshift clusters would allow us to constrain the relationship between weak lensing mass and directly observable quantities in the local universe, particularly for subsets of cluster populations, e.g., relaxed vs. unrelaxed systems. Our efforts could supplement the substantial progress made in the $z\gtrsim 0.2$ universe by~\cite{2014HEAD...1411108M},~\cite{2016MNRAS.456L..74S} and~\cite{2016MNRAS.457.1522A}, among others.

Three of the four clusters in the sample (A2029, A1606 and A85) show small ($\sim 2-3\arcmin$, 100-250 kpc) but noticeable offsets between the peak of the WL signal and the brightest cluster galaxy, cf. Figures~\ref{fig:A85sigma} to~\ref{fig:A1606overlay}. These offsets are likely genuine: the simulations of Section~\ref{subsec:nfwtests} show that the uncertainty in the WL centroid from random shape noise is small ($1\farcm1$ or less) for clusters in this mass range. The relative distributions of cluster dark matter, X-rays, and BCGs can place strong limits on potential dark matter models (e.g.~\citealp{2018MNRAS.477..669M}), and expanding our weak lensing analysis to more clusters could contribute to this effort.

\acknowledgments

The research was carried out at the Jet Propulsion Laboratory, California Institute of Technology, under a contract with the National Aeronautics and Space Administration. JM gratefully acknowledges the use of data reduction and PSF correction code developed by Thomas Erben and Mark Allen. JM also gratefully acknowledges useful discussions with Steve Allen, Douglas Applegate and Adam Wright over the course of this project, particularly regarding the implementation of data reduction codes. \textcopyright~2018 All Rights Reserved.
\vspace{5mm}
\facilities{CTIO:4m (DECam)}


\software{astropy \citep{2013A&A...558A..33A}, 
          SExtractor \citep{1996A&AS..117..393B},
          SWarp \citep{2010ascl.soft10068B},
          Seaborn (doi:10.5281/zenodo.883859)
          }


\begin{thebibliography}{}
\bibitem[Agulli et al.(2016)]{2016MNRAS.458.1590A} Agulli, I., Aguerri, J.~A.~L., S{\'a}nchez-Janssen, R., et al.\ 2016, \mnras, 458, 1590 
\bibitem[Applegate et al.(2014)]{2014MNRAS.439...48A} Applegate, D.~E., von der Linden, A., Kelly, P.~L., et al.\ 2014, \mnras, 439, 48 \bibitem[Astropy Collaboration et al.(2013)]{2013A&A...558A..33A} Astropy Collaboration, Robitaille, T.~P., Tollerud, E.~J., et al.\ 2013, \aap, 558, A33 
\bibitem[Applegate et al.(2016)]{2016MNRAS.457.1522A} Applegate, D.~E., Mantz, A., Allen, S.~W., et al.\ 2016, \mnras, 457, 1522.
\bibitem[\protect\astroncite{Bartelmann \& Schneider}{2001}]{2001PhR...340..291B} Bartelmann, M., \& Schneider, P.\ 2001, \physrep, 340, 291 
\bibitem[Ben{\'{\i}}tez (2000)]{2000ApJ...536..571B} Ben{\'{\i}}tez, N.\ 2000, \apj, 536, 571 
%\bibitem[Bernstein \& Jarvis (2002)]{2002AJ....123..583B} Bernstein, G.~M., \& Jarvis, M.\ 2002, \aj, 123, 583 
\bibitem[Bertin \& Arnouts (1996)]{1996A&AS..117..393B} Bertin, E., \& Arnouts, S.\ 1996, \aaps, 117, 393 
%\bibitem[Bertin et al.(2002)]{2002ASPC..281..228B} Bertin, E., Mellier, Y., Radovich, M., et al.\ 2002, Astronomical Data Analysis Software and Systems XI, 281, 228 
\bibitem[Bertin (2010)]{2010ascl.soft10068B} Bertin, E.\ 2010, Astrophysics Source Code Library, ascl:1010.068 
\bibitem[Bhattacharya et al.(2013)]{2013ApJ...766...32B} Bhattacharya, S., Habib, S., Heitmann, K., \& Vikhlinin, A.\ 2013, \apj, 766, 32 
\bibitem[Coe et al.(2006)]{2006AJ....132..926C} Coe, D., Ben{\'{\i}}tez, N., S{\'a}nchez, S.~F., et al.\ 2006, \aj, 132, 926 
%\bibitem[Diemand et al.(2007)]{2007ApJ...667..859D} Diemand, J., Kuhlen, M., \& Madau, P.\ 2007, \apj, 667, 859 
%\bibitem[Diemer \& Kravtsov(2015)]{2015ApJ...799..108D} Diemer, B., \& Kravtsov, A.~V.\ 2015, \apj, 799, 108 
\bibitem[DePoy et al.(2008)]{2008SPIE.7014E..0ED} DePoy, D.~L., Abbott, T., Annis, J., et al.\ 2008, \procspie, 7014, 70140E 
\bibitem[Durret et al.(2005)]{2005A&A...432..809D} Durret, F., Lima Neto, G.~B., \& Forman, W.\ 2005, \aap, 432, 809 
\bibitem[Erben et al.(2001)]{2001A&A...366..717E} Erben, T., Van Waerbeke, L., Bertin, E., Mellier, Y., \& Schneider, P.\ 2001, \aap, 366, 717 
%\bibitem[Fischer \& Tyson (1997)]{1997AJ....114...14F} Fischer, P., \& Tyson, J.~A.\ 1997, \aj, 114, 14 
\bibitem[Flaugher et al.(2015)]{2015AJ....150..150F} Flaugher, B., Diehl, H.~T., Honscheid, K., et al.\ 2015, \aj, 150, 150 
%\bibitem[Gao et al.(2004)]{2004MNRAS.355..819G} Gao, L., White, S.~D.~M., Jenkins, A., Stoehr, F., \& Springel, V.\ 2004, \mnras, 355, 819 
%\bibitem[Gao et al.(2012)]{2012MNRAS.425.2169G} Gao, L., Navarro, J.~F., Frenk, C.~S., et al.\ 2012, \mnras, 425, 2169
\bibitem[Geller et al.(2010)]{2010ApJ...709..832G} Geller, M.~J., Kurtz, M.~J., Dell'Antonio, I.~P., Ramella, M., \& Fabricant, D.~G.\ 2010, \apj, 709, 832 
%\bibitem[Gonzalez et al.(2013)]{2013ApJ...778...14G} Gonzalez, A.~H., Sivanandam, S., Zabludoff, A.~I., \& Zaritsky, D.\ 2013, \apj, 778, 14 
\bibitem[Gruen et al.(2014)]{2014PASP..126..158G} Gruen, D., Seitz, S., \& Bernstein, G.~M.\ 2014, \pasp, 126, 158 
\bibitem[Gullieuszik et al.(2015)]{2015A&A...581A..41G} Gullieuszik, M., Poggianti, B., Fasano, G., et al.\ 2015, \aap, 581, A41 
\bibitem[Harvey et al.(2019)]{2019MNRAS.488.1572H} Harvey, D., Robertson, A., Massey, R., et al.\ 2019, \mnras, 488, 1572
%\bibitem[Heitmann et al.(2014)]{2014ApJ...780..111H} Heitmann, K., Lawrence, E., Kwan, J., Habib, S., \& Higdon, D.\ 2014, \apj, 780, 111 
\bibitem[Herbonnet et al.(2019)]{2019MNRAS.490.4889H} Herbonnet, R., von der Linden, A., Allen, S.~W., et al.\ 2019, \mnras, 490, 4889
\bibitem[\protect\astroncite{Hetterscheidt et al.}{2005}]{2005A&A...442...43H} Hetterscheidt, M., Erben, T., Schneider, P., et al.\ 2005, \aap, 442, 43 
\bibitem[Heymans et al.(2006)]{2006MNRAS.368.1323H} Heymans, C., Van Waerbeke, L., Bacon, D., et al.\ 2006, \mnras, 368, 1323 
\bibitem[Hoekstra et al.(1998)]{1998ApJ...504..636H} Hoekstra, H., Franx, M., Kuijken, K., \& Squires, G.\ 1998, \apj, 504, 636 
\bibitem[Hu \& Kravtsov(2003)]{2003ApJ...584..702H} Hu, W., \& Kravtsov, A.~V.\ 2003, \apj, 584, 702.
\bibitem[\protect\astroncite{Huwe}{2013}]{huwethesis}{Huwe}, P.~M. 2013, Dark Matter Substructure in High Redshift Clusters of Galaxies, Ph.D. Thesis, Brown University 
\bibitem[Ichinohe et al.(2015)]{2015MNRAS.448.2971I} Ichinohe, Y., Werner, N., Simionescu, A., et al.\ 2015, \mnras, 448, 2971 
\bibitem[\protect\astroncite{Jarvis et al.}{2003}]{2003AJ....125.1014J} Jarvis, M., Bernstein, G.~M., Fischer, P., et al.\ 2003, \aj, 125, 1014 
\bibitem[Kaiser et al.(1995)]{1995ApJ...449..460K} Kaiser, N., Squires, G., \& Broadhurst, T.\ 1995, \apj, 449, 460 
\bibitem[Kelly et al.(2014)]{2014MNRAS.439...28K} Kelly, P.~L., von der Linden, A., Applegate, D.~E., et al.\ 2014, \mnras, 439, 28 
\bibitem[Kempner et al.(2002)]{2002ApJ...579..236K} Kempner, J.~C., Sarazin, C.~L., \& Ricker, P.~M.\ 2002, \apj, 579, 236 
\bibitem[Limousin et al.(2009)]{2009ApJ...696.1771L} Limousin, M., Sommer-Larsen, J., Natarajan, P., \& Milvang-Jensen, B.\ 2009, \apj, 696, 1771 
\bibitem[Lakhchaura \& Singh(2014)]{2014AJ....147..156L} Lakhchaura, K., \& Singh, K.~P.\ 2014, \aj, 147, 156 
\bibitem[Luppino \& Kaiser (1997)]{1997ApJ...475...20L} Luppino, G.~A., \& Kaiser, N.\ 1997, \apj, 475, 20 
\bibitem[Massey et al.(2007)]{2007MNRAS.376...13M} Massey, R., Heymans, C., Berg{\'e}, J., et al.\ 2007, \mnras, 376, 13 
\bibitem[Massey et al.(2018)]{2018MNRAS.477..669M} Massey, R., Harvey, D., Liesenborgs, J., et al.\ 2018, \mnras, 477, 669
\bibitem[McCleary et al.(2015)]{2015ApJ...805...40M} McCleary, J., dell'Antonio, I., \& Huwe, P.\ 2015, \apj, 805, 40
%\bibitem[Morandi et al.(2012)]{2012MNRAS.425.2069M} Morandi, A., Limousin, M., Sayers, J., et al.\ 2012, \mnras, 425, 2069 
\bibitem[Mahdavi et al.(2008)]{2008MNRAS.384.1567M} Mahdavi, A., Hoekstra, H., Babul, A., et al.\ 2008, \mnras, 384, 1567
\bibitem[Mahdavi et al.(2014)]{2014HEAD...1411108M} Mahdavi, A., Hoekstra, H., \& Babul, A.\ 2014, AAS/High Energy Astrophysics Division \#14 , 111.08.
%\bibitem[Navarro et al.(1997)]{1997ApJ...490..493N} Navarro, J.~F., Frenk, C.~S., \& White, S.~D.~M.\ 1997, \apj, 490, 493  
\bibitem[Oyaizu et al.(2008)]{2008ApJ...689..709O} Oyaizu, H., Lima, M., Cunha, C.~E., et al.\ 2008, \apj, 689, 709
\bibitem[Paterno-Mahler et al.(2013)]{2013ApJ...773..114P} Paterno-Mahler, R., Blanton, E.~L., Randall, S.~W., \& Clarke, T.~E.\ 2013, \apj, 773, 114 
\bibitem[Piffaretti et al.(2011)]{2011A&A...534A.109P} Piffaretti, R., Arnaud, M., Pratt, G.~W., Pointecouteau, E., \& Melin, J.-B.\ 2011, \aap, 534, A109 
%\bibitem[Pillepich et al.(2018)]{2018MNRAS.481..613P} Pillepich, A., Reiprich, T.~H., Porciani, C., et al.\ 2018, \mnras, 481, 613
\bibitem[Polletta et al.(2007)]{2007ApJ...663...81P} Polletta, M., Tajer, M., Maraschi, L., et al.\ 2007, \apj, 663, 81 
\bibitem[\protect\astroncite{Schirmer et al.}{2004}]{2004A&A...420...75S} Schirmer, M., Erben, T., Schneider, P., Wolf, C., \& Meisenheimer, K.\ 2004, \aap, 420, 75 
\bibitem[\protect\astroncite{Schneider}{1996}]{1996MNRAS.283..837S} Schneider, P.\ 1996, \mnras, 283, 837 
\bibitem[Schuecker et al.(2001)]{2001A&A...378..408S} Schuecker, P., B{\"o}hringer, H., Reiprich, T.~H., \& Feretti, L.\ 2001, \aap, 378, 408 
\bibitem[Shaw~(2015)]{NOAODHB2.2}Shaw, R.~A., (ed.) 2015, NOAO Data Handbook (Version 2.2; Tucson, AZ: National Optical Astronomy Observatory)
\bibitem[Smith et al.(2016)]{2016MNRAS.456L..74S} Smith, G.~P., Mazzotta, P., Okabe, N., et al.\ 2016, \mnras, 456, L74.
\bibitem[Sohn et al.(2018)]{2018arXiv180901137S} Sohn, J., Geller, M.~J., Zahid, H.~J., et al.\ 2018, ArXiv e-prints , arXiv:1809.01137.
\bibitem[Tucker et al.(2000)]{2000ApJS..130..237T} Tucker, D.~L., Oemler, A., Jr., Hashimoto, Y., et al.\ 2000, \apjs, 130, 237 
%\bibitem[von der Linden et al.(2014)]{2014MNRAS.439....2V} von der Linden, A., Allen, M.~T., Applegate, D.~E., et al.\ 2014, \mnras, 439, 2 
\bibitem[Walker et al.(2012)]{2012MNRAS.422.3503W} Walker, S.~A., Fabian, A.~C., Sanders, J.~S., George, M.~R., \& Tawara, Y.\ 2012, \mnras, 422, 3503 
\bibitem[\protect\astroncite{Wittman}{2002a}]{2002LNP...608...55W} {Wittman}, D.\ 2002, in Lecture Notes in Physics, ed. {Courbin}, F. and {Minniti}, D. (Berlin: Springer-Verlink), 608, 55 
\bibitem[Wang et al.(2014)]{2014MNRAS.445..614W} Wang, M.-Y., Peter, A.~H.~G., Strigari, L.~E., et al.\ 2014, \mnras, 445, 614
\bibitem[Wright \& Brainerd(2000)]{2000ApJ...534...34W} Wright, C.~O., \& Brainerd, T.~G.\ 2000, \apj, 534, 34
%\bibitem[Yu et al.(2016)]{2016ApJ...831..156Y} Yu, H., Diaferio, A., Agulli, I., et al.\ 2016, \apj, 831, 156.

\end{thebibliography}
\end{document}